\documentclass[acmsmall,screen,nonacm]{acmart}
\usepackage[T1]{fontenc}
\usepackage{silence}
\AtBeginDocument{%
  }
    
\usepackage{amsmath,amsfonts}
\usepackage{mathtools}
\usepackage{xspace}
\usepackage{booktabs}
\usepackage{xcolor}
\usepackage{listings}
\usepackage{enumitem}
\usepackage{cleveref}
\usepackage{microtype}
\usepackage{algorithm}
\usepackage{algpseudocode}
\usepackage{physics}
\usepackage{amsthm}
\usepackage{textcomp}
\usepackage{graphicx}
\usepackage{hyperref}
\usepackage{wrapfig}
\newcommand{\PGCconfigs}{48}
\newcommand{\PGCcertMinCov}{1.000}
\newcommand{\PGCslackMin}{4.8}
\newcommand{\PGCslackMax}{6.7}
\newcommand{\PGCentryFail}{24}
\newcommand{\PGCentryTotal}{24}
\newcommand{\PGCentryMinCov}{0.00}
\newcommand{\PGCentryMaxCov}{0.03}
\newcommand{\PGCgammaRatio}{0.896}
\newcommand{\PGCsessionsOne}{1000}
\newcommand{\PGCsessionsTwo}{400}
\newcommand{\PGCgammaN}{20}
\newcommand{\PGCepsCertMin}{0.21}
\newcommand{\PGCepsCertMax}{1.79}
\newcommand{\PGCepsCertMinTmax}{0.21}
\newcommand{\PGCepsCertMaxTmax}{0.45}
\newcommand{\PGCholdoutFive}{0.957}
\newcommand{\PGCholdoutOne}{0.990}

\newcommand{\ReachConf}{0.95}

\newcommand{\RQthreeBeta}{0.33}
\newcommand{\RQthreeBetaSd}{0.05}
\newcommand{\RQthreeBetaMin}{0.23}
\newcommand{\RQthreeBetaMax}{0.38}
\newcommand{\RQthreeBetaMed}{0.42}
\newcommand{\RQthreeBetaMedSd}{0.08}
\newcommand{\RQthreeXgrowth}{6.7}
\newcommand{\RQthreeTgrowthMin}{1.6}
\newcommand{\RQthreeTgrowthMax}{2.0}
\newcommand{\RQthreeTmax}{5{,}366}
\newcommand{\RQthreeTmaxMed}{2{,}369}
\newcommand{\RQfourCold}{3{,}469}

\newcommand{\RQfourWarm}{1{,}539}
\newcommand{\RQfourWarmSd}{167}
\newcommand{\RQfourRatio}{2.3}
\newcommand{\RQfourColdMed}{1{,}480}
\newcommand{\RQfourWarmMed}{54}
\newcommand{\RQfourRatioMed}{27}
\newcommand{\RQfourDrift}{0.050}
\newcommand{\ScaleBeta}{0.28}

\newcommand{\ScaleXmin}{8}
\newcommand{\ScaleXmax}{304}
\newcommand{\ScaleXrange}{38}
\newcommand{\ScaleTfirst}{3{,}153}
\newcommand{\ScaleTlast}{8{,}290}
\newcommand{\ScaleBetaWithinMin}{0.08}
\newcommand{\ScaleBetaWithinMax}{0.25}
\newcommand{\LoneExpo}{2.06}
\newcommand{\LoneTdepthHundred}{5.4\times10^{4}}
\newcommand{\LoneFdepthHundred}{4.8\times10^{-2}}
\newcommand{\LoneTdepthTwentyFive}{9.0\times10^{5}}
\newcommand{\LoneFsingle}{4.9\times10^{-4}}
\newcommand{\LoneTsingle}{7\times10^{8}}

\newcommand{\LfourA}{2.0}

\newcommand{\LfourB}{1.4}

\newcommand{\LfourC}{0.3}

\newcommand{\LfourCross}{0.16}
\newcommand{\ShotsM}{2,4,8,12}
\newcommand{\ShotsT}{800,800,800,1600}
\newcommand{\ShotsTmed}{200,400,400,400}
\newcommand{\ShotsEps}{0.20}
\newcommand{\LthreeDiamond}{0.0396}

\newcommand{\GCpauliChannels}{82}
\newcommand{\GCpauliPairs}{6{,}724}
\newcommand{\GCpauliViolations}{0}
\newcommand{\GCpauliOrdered}{1{,}691}
\newcommand{\GCnonPauliChannels}{43}
\newcommand{\GCnonPauliPairs}{1{,}849}

\newcommand{\GCnonPauliDistinctOrdered}{0}

\newcommand{\Var}{\operatorname{Var}}

\newcommand{\States}{\mathcal{S}}     
\newcommand{\Obs}{\mathcal{O}}         

\definecolor{codegreen}{rgb}{0,0.6,0}
\definecolor{codegray}{rgb}{0.5,0.5,0.5}
\definecolor{codepurple}{rgb}{0.58,0,0.82}
\definecolor{backcolour}{rgb}{0.95,0.95,0.92}
\lstdefinestyle{mystyle}{
    backgroundcolor=\color{backcolour},   
    commentstyle=\color{codegreen},
    keywordstyle=\color{magenta},
    numberstyle=\tiny\color{codegray},
    stringstyle=\color{codepurple},
    basicstyle=\ttfamily\footnotesize,
    breakatwhitespace=false,         
    breaklines=true,                 
    captionpos=b,                    
    keepspaces=true,                 
    numbers=left,                    
    numbersep=5pt,                  
    showspaces=false,                
    showstringspaces=false,
    showtabs=false,                  
    tabsize=2
}
\newcommand{\mypara}[1]{%
  \par\parskip=2pt
  \noindent\textit{#1. }\ignorespaces
}

\usepackage{enumitem} 
\setlist[itemize]{noitemsep, nolistsep}

\newcommand{\Rgate}{$R_z(\theta) = \begin{bmatrix} e^{-i\frac{\theta}{2}} & 0 \\ 0 & e^{i\frac{\theta}{2}} \end{bmatrix}$}

\newcommand{\Rygate}{$R_y(\theta) = \begin{bmatrix} \cos\left(\theta/2\right) & -\sin\left(\theta/2\right) \\ \sin\left(\theta/2\right) & \cos\left(\theta/2\right) \end{bmatrix}$}

\newcommand{\Rxgate}{$R_x(\theta) =  \begin{bmatrix} \cos\left(\theta/2\right) & -i\sin\left(\theta/2\right) \\ -i\sin\left(\theta/2\right) & \cos\left(\theta/2\right) \end{bmatrix}$}

\newcommand*{\eg}{e.g.\@\xspace}
\newcommand*{\ie}{i.e.\@\xspace}

\newcommand{\code}[1]{\texttt{#1}}

\crefformat{section}{§#2#1#3}
\Crefformat{section}{§#2#1#3}

\newcommand{\Chan}{\mathcal{C}}
\theoremstyle{plain}
\newtheorem{definition}{Definition}
\newtheorem{proposition}{Proposition}
\newtheorem{theorem}{Theorem}

\newcommand{\tool}{\textsc{ChanShadow}\xspace}

\title{Live Noise Fingerprints for Quantum Program Analysis}

\author{Avner Bensoussan}
\orcid{0009-0007-3285-9468}
\affiliation{
    \institution{King's College London}
    \city{London}
    \country{United Kingdom}
}
\email{avner.bensoussan@kcl.ac.uk}

\author{Karine Even-Mendoza}
\orcid{0000-0002-3099-1189}
\affiliation{
    \institution{King's College London}
    \city{London}
    \country{United Kingdom}
}

\author{Debalina Banerjee}
\orcid{0000-0003-3943-5389}
\affiliation{
    \institution{University of Oxford}
    \city{Oxford}
    \country{United Kingdom}
}

\author{Elena Chachkarova}
\orcid{0000-0003-2857-5570}
\affiliation{
    \institution{King's College London}
    \city{London}
    \country{United Kingdom}
}

\author{Sophie Fortz}
\orcid{0000-0001-9687-8587}
\affiliation{
    \institution{Inria, Univ. Rennes, CNRS, IRISA}
    \city{Rennes}
    \country{France}
}

\author{Vasileios Klimis}
\orcid{0000-0002-3173-8636}
\affiliation{
    \institution{Queen Mary University of London}
    \city{London}
    \country{United Kingdom}
}

\author{Mohammad Reza Mousavi}
\orcid{0000-0002-4869-6794}
\affiliation{
    \institution{King's College London}
    \city{London}
    \country{United Kingdom}
}

\begin{document}

\begin{abstract}

Quantum programs are probabilistic in nature, and noisy execution due to current imperfect hardware further complicates their analysis. In this paper, we study abstract semantics for analysing quantum programs under these conditions. We introduce \emph{noise fingerprints}, compact descriptions of the observable effects of noise relative to a finite interface of input states and output measurements. We use them to define our abstract domain with respect to the concrete noise-related behaviour. We relate the concrete and abstract domains through a \emph{probabilistic Galois connection}, extending the Galois-connection framework to distribution-valued interpretations.

We study the resulting abstraction in terms of observational soundness, conciseness, efficiency, and ability to adapt to changing noise. 
We show that polynomial-size interfaces yield exponentially smaller fingerprints than full noise models while preserving the chosen observations, and that fingerprint distance characterises the largest observational discrepancy over the chosen interface. 
Further, we develop \tool{}, which reuses noise fingerprints of commonly used short program fragments to analyse noisy execution of whole programs and provides a uniform finite-sample error bound. We construct working examples showing how fingerprints support program analysis.
We evaluate fingerprint precision, representation cost, measurement requirements, and behaviour under noise drift. On a synthetic benchmark, fingerprints retain $(99 \pm 1)\%$ of the noise strength information of a full noise model. 
On simulated noise drift, reusing earlier measurements reduces the number of fresh measurement rounds needed to reach an error tolerance of $0.10$ by a factor of $2.3$ at $95\%$ confidence and by a factor of $27$ under the median-error criterion.

\end{abstract}

\maketitle

\section{Introduction}
\label{sec:intro}
Quantum software engineering and development 
rely on noisy quantum computer infrastructures~\cite{bharti_noisy_2022,10.1145/3297858.3304075,10123493}. 
Noisy executions pose a challenge to quantum program testing 
by making failures harder to attribute~\cite{10682972} and limiting the applicability of existing testing methods~\cite{Wang2025:thelandscape}. For example, noise can change which faults a test suite appears to detect in mutation testing~\cite{fortz2026robust} and lead to misjudgements about program behaviour in measurement-based validation~\cite{11334299}.
The implications extend beyond testing: noise can itself cause incorrect program behaviour~\cite{11029802,10.1145/3689716,10.1145/3290344}.

Consequently, a growing body of work on testing, debugging, and program analysis recognises noise as a challenge~\cite{fuzzq,Chareton:2021age,huang2025efficient,Maschek_2025,QSERoadmap2025,10.1145/3624483}, 
explores machine learning to mitigate its effects~\cite{10682972}, considers statevector-based validation~\cite{11334299}, evaluates comparison metrics under noise~\cite{fortz2026robust}, and develops analyses that explicitly account for noise~\cite{simshadow:ICST:2026,muroya2026noiseawareverificationsynthesisquantum}. 
Further, developers already account for noise when designing algorithms, requiring development environments that support simulation and analysis of noisy execution \cite{10.1145/3838727,Maschek_2025,ibmquantumbuildnoisemodels}:
The noise behaviour observed at this stage feeds directly into code development~\cite{10123493,10682972} and into optimisation and transpilation decisions~\cite{10.1145/3428198,10.1145/3297858.3304075,10821340}.

Nonetheless, when noise changes, existing techniques can require substantial rework: recalibrating detection thresholds~\cite{fortz2026robust}, adapting learned models to the device and circuit~\cite{10682972}, or repeating verification with an updated noise model~\cite{muroya2026noiseawareverificationsynthesisquantum}. This exposes a real difficulty as the effective noise of a noisy backend is opaque, drifts---sometimes within hours~\cite{martina2022learning}---and differs across platforms. 
Even when two frameworks expose the \emph{same} named noise model with the \emph{same} nominal parameters, their effective behaviour diverges due to implementation details, compilation passes, numerical precision, and backend-specific modelling details~\cite{fuzzq,simshadow:ICST:2026,10.21468/SciPostPhysCore.7.4.075} (\eg{} The results in \cite{simshadow:ICST:2026} showed non-negligible divergence of \emph{same} programs \cite{quetschlich2023mqtbench} under the \emph{same} circuit representation between Qiskit and Cirq under the \emph{same} IBM Boston noise profile).

To reason about these differences, we need a model of how noisy execution environments affect the program. One such model represents a small phase error caused by noise as an extra gate in the program’s gate sequence~\cite{10.1145/3689716}. For example, the ideal program $P = H(q_0); CX(q_0,q_1)$ can become in noisy execution environment 
\begin{equation}\label{eq:noisyprogram}
P_{\mathrm{noisy}} = H(q_0); R_z(0.30)(q_0); CX(q_0,q_1).  
\end{equation}
Whether the additional phase rotation of $0.30$ radians error is detectable 
depends on how noise is modelled and which properties of program behaviour the analysis considers, as we discuss in~\cref{sec:motivation}.
Eq. \ref{eq:noisyprogram} example isolates a specific effect of noise. 

In this paper, we suggest modelling noise in the execution environment as a \textit{quantum channel},
a mathematical map describing how noise transforms a quantum state, not just a noise phase rotation. This reflects current practice in quantum software development, where practitioners reason about noise through its effects on \emph{observed} program behaviour, such as changes in measured expectation values and their statistics~\cite{huang2020predicting,10682972}---rather than internal noise mechanisms~\cite{PhysRevA.65.062101,Knapp2018modelingnoiseerror}.
Yet a full description of a noisy quantum channel grows exponentially with the number of qubits \cite{levy2024shadowqpt,simshadow:ICST:2026}, making it impractical to compute, store, or compare during routine development, even when the analysis concerns only a few observations.
A developer often asks a more specific question. \textit{Did a change affect the outputs used by a test or an intermediate behaviour that helps explain a final failure?} 
Adding to this, a change in noise can require rebuilding the model or rerunning the analysis from scratch. What is missing is a sound, compact abstraction that captures the channel’s effects on the relevant observations, is inexpensive to obtain, and can be updated as noise changes.

\parskip=2pt

\mypara{\textbf{Our Contribution: An abstract domain for noisy channels}}
We study the semantic problem of constructing tractable abstractions of noisy quantum channels:
\textit{Given a finite set of observations, how can we summarise channel behaviour while preserving properties of interest?}

We model noisy quantum executions using quantum channels, \ie{} \textit{Completely Positive Trace-Preserving} (CPTP) maps from input states to output states. A pure state can be written as a ket vector $\lvert x\rangle$. A density matrix $\rho$ can represent a pure or a mixed state. We select observations before characterising the channel.
Each \emph{observation} pairs a prepared input with a property measured at the output. The resulting finite table is an \emph{observation profile}, or \emph{noise fingerprint}. Thus, environments with similar entries behave similarly on the experiments of these entries and may differ on other inputs or output measurements. This is the core idea behind our abstraction of a quantum channel: an \emph{interface-relative noise fingerprints} as an abstract semantics for noisy quantum execution, leveraging the idea that reasoning about and comparing noisy executions often requires only the observations relevant to the properties under study. Nonetheless, the selected observations determine the fingerprint’s scope, and the sample size determines the analysis’s accuracy. Building an abstract model for each property of interest addresses a gap in existing methods, reducing analysis reruns as code changes and noise drifts.

This is, at its core, a problem of abstract interpretation~\cite{Cousot77,wilhelm_principles_2022}. We cast the effective noise of a channel as an \emph{abstract domain} whose elements are \emph{noise fingerprints}: finite abstractions of channel behaviour defined by an experiment interface $X$.
The abstraction \(\alpha_X\) maps a concrete channel \(\Lambda\) to its exact fingerprint on interface \(X\). Probabilistic concretisation \(\gamma_P\) maps an abstract fingerprint \(F\) to a probability distribution over concrete channels. Its expected channel is \(\bar{\gamma}_P(F)\).
Together, $\alpha$ and $\gamma_P$ form a probabilistic Galois connection (\cref{def:galois}) that reduces to the classical one when every concretisation is a point mass. 
Separately, a finite-shot protocol induces a distribution $\alpha_T$ over measured fingerprints\footnote{Each run returns one estimate of the exact fingerprint \(\alpha_X(\Lambda)\).}, whose $(\varepsilon,\delta)$ guarantee is part of estimation efficiency (\cref{def:est-eff}).

We study four properties of the abstraction that guide the design (\cref{sec:core}) and evaluation (\cref{sec:eval}) of our framework:
(i)~observational~\emph{soundness}, that the measured abstraction and tolerance-based concretisation satisfy a probabilistic coverage guarantee and that observable behaviour is stable under changes in the abstract domain; 
(ii)~\emph{conciseness}, that the abstraction compresses the exponential concrete domain objects while retaining the observable information of interest;
(iii)~estimation~\emph{efficiency}, that fingerprints are cheap to construct and to use; and (iv)~\emph{adaptiveness}, that they respond continuously and incrementally to changes in the underlying noise across platforms, configurations, and time.
The framework admits arbitrary experiment interfaces. In our evaluation, we instantiate the framework with the interface $X = \States \times \Obs$, consisting of reference states ($\States$) and observables ($\Obs$), where a fingerprint records the corresponding expectation-value deviations. This is one instantiation, built on the results of SimShadow \cite{simshadow:ICST:2026}.

\parskip=2pt
To summarise, the main contributions of the paper are:
\begin{itemize} [leftmargin=*,noitemsep,topsep=0pt]
    \item \textbf{Program-level working examples. }%
    We illustrate our analysis on a simple program (\cref{sec:motivation}). In Appendix \cref{app:diff-debug}, we motivate using fingerprints to check compiler optimisations, assess test portability, and support debugging. Our artifact includes a Jupyter notebook for reproducibility, accepting QASM programs as input.

    \item \textbf{Fingerprints as an abstract domain. }%
    We model noise fingerprints as an observational abstraction of CPTP channels over a finite
    interface (\cref{sec:properties}).
    
    \item \textbf{Extending the notion of Galois connection to a probabilistic setting. }%
    We introduce \emph{probabilistic Galois connections} to randomized, empirically estimated abstractions (\cref{sec:properties}). 

    \item \textbf{A classical-shadow construction (\tool). }%
    We instantiate the abstract domain with $\alpha_{\mathrm{shadow}}$, a parallel classical-shadow estimator with a logarithmic sample-complexity guarantee, and a non-parametric semidefinite program concretisation $\gamma_{\mathrm{shadow}}$ (\cref{sec:concrete}). 


  \item \textbf{Evaluation against theoretical bounds. }%
  We establish exponential compression and (poly)loga- rithmic sample complexity, and observe tight soundness bounds for amplitude and phase damping and incremental convergence at the predicted rate in simulation (\cref{sec:eval}).

  \item \textbf{Expert verification and identifiability. }%
  Threats to Validity and limitations are discussed in \cref{sec:appNdiscuss}. We verify that fingerprints reproduce the physically correct per-noise signatures and can separate noise families and expose real cross-platform discrepancies (Appendix, \cref{sec:theory2Emp}). 

\end{itemize}
\section{Background and Motivation}
\label{sec:mot}
\label{sec:motivation}
\label{sec:example}
\label{sec:vignette}

We briefly review quantum computing basics~\cite{nielsen2010quantum}, with further details in \cref{subsec:background}. A pure quantum state is a unit vector $\lvert\psi\rangle$ in a complex Hilbert space, with dual $\langle\psi\rvert=\lvert\psi\rangle^\dagger$. 
Gates such as $H$ and $CX$ are unitary operators. An $n$-qubit (mixed) state is represented by a density matrix $\rho\in\mathbb{C}^{2^n\times 2^n}$, with $\rho=\lvert\psi\rangle\langle\psi\rvert$ in the pure case. 
An \textit{observable} $O$ is a Hermitian operator whose eigenvalues are the possible measurement outcomes. The expected result of measuring $O$ in state $\rho$ is $\operatorname{Tr}(O\rho)$, which equals $\langle\psi\rvert O\lvert\psi\rangle$ for a pure state $\lvert\psi\rangle$.
A quantum channel $\Lambda$ is a function from input density matrices to output density matrices; \eg{} Qiskit’s 
\href{https://docs.quantum.ibm.com/api/qiskit/qiskit.quantum_info.Kraus}{Kraus documentation} gives the function explicitly (see also Appendix \cref{sec:apx:shots}).
For example, the Kraus matrices for \href{https://qiskit.github.io/qiskit-aer/stubs/qiskit_aer.noise.phase_damping_error.html}{Qiskit Aer phase-damping} give the following single-qubit function, where $p$ is the phase-damping parameter:
\[
\Lambda_p\!\left(
\begin{pmatrix}a&b\\c&d\end{pmatrix}
\right)
=
\begin{pmatrix}
a & \sqrt{1-p}\,b\\
\sqrt{1-p}\,c & d
\end{pmatrix}.\]%

\paragraph{Example of Testing under Noise}
Consider the program $P$ above
executed from the initial state $\lvert 00\rangle$. Ideally, the program prepares a Bell state, in which the two qubits are correlated. 
We use three test assertions, one for each of the Pauli observables $ZZ$, $XX$, and $YY$. Together, they measure output correlations in the computational and superposition bases.

Now consider two execution environments for the same program. Environment A is a baseline configuration, with depolarising noise and amplitude damping after each gate. Environment B has the same noise settings but adds the phase rotation as in $P_{\mathrm{noisy}}$ (eq. \ref{eq:noisyprogram}) to the baseline noise model.
%
To compare the two executions, consider how much each measured property deviates from its ideal value. For the input $\lvert 00\rangle$, these deviations are:   
\[
\begin{array}{c|rrr}
 & ZZ & XX & YY\\
\hline
\text{Environment A} & -0.1152 & -0.1288 &  0.1288\\
\text{Environment B} & -0.1152 & -0.1677 &  0.1677
\end{array}
\]%
The two environments behave identically with respect to $ZZ$, but differ by $0.0389$ on both $XX$ and $YY$. With a comparison tolerance of $\varepsilon=0.02$, the difference is therefore observable. The phase rotation leaves the population check unchanged but affects the correlations measured in the superposition bases.

The choice of checks matters. Suppose the same program is tested using only population measurements, such as $ZZ$, $ZI$, and $IZ$. The two environments then have a distance of $0.0000$ on these checks and would be considered equivalent. This does not mean that their behaviour is identical: the selected measurements simply do not expose the difference. The same environmental change can therefore be relevant to one test and invisible to another.



This example illustrates how our framework supports comparisons between execution environments. Detailed examples with further scenarios are available in~\Cref{app:diff-debug} and the \href{https://zenodo.org/records/23414157}{replication package}.

\section{Interface-Relative Semantics and Abstraction Properties}
\label{sec:core}
\label{sec:properties}
\label{sec:semantics}


This section gives noise fingerprints a precise semantic meaning by describing a channel through a finite interface of experiments. 
The resulting fingerprint is exact with respect to that interface but makes no claim about behaviour the interface cannot observe. We define the interface and exact fingerprint, relate abstract and concrete domains via an expectation-based probabilistic Galois connection, and formulate four properties: soundness, conciseness, estimation efficiency, and adaptiveness, guiding the construction and evaluation throughout the paper.

\subsection{Experiment Interfaces and Exact Fingerprints}
\label{subsec:interface-semantics}

\begin{definition}[Quantum Channel]
A quantum channel on $n$ qubits is a completely positive,
trace-preserving (CPTP) linear map on density matrices.
We write $\mathcal{C}_n$ for the set of such channels.
Further details, including the Kraus representation, are given in Appendix \cref{sec:apx:shots}.
\end{definition}

An experiment prepares a reference state, applies the channel under study, and measures an observable. An analysis chooses in advance which such experiments are relevant.

\begin{definition}[Experiment Interface]\label{def:reference}
Let $\States=\{\rho_1,\ldots,\rho_k\}$ be reference density
operators on the $n$-qubit space and let $\Obs=\{O_1,\ldots,O_m\}$ be
Hermitian observables with $\|O_j\|_{\mathrm{op}}\leq 1$.
The experiment interface is $X=\States\times\Obs$.
A fingerprint over $X$ is a function $F:X\to\mathbb{R}$
(equivalently, a $k\times m$ matrix), with one entry
$F(\rho,O)$ for each experiment that prepares $\rho$,
applies the channel, and measures $O$.
\end{definition}

\begin{definition}[Exact Interface Fingerprint]\label{def:exact-fingerprint}
For $\Lambda\in\Chan$, define its exact fingerprint by
\[
  \alpha_X(\Lambda)(\rho,O)
  =\Tr[O\Lambda(\rho)]-\Tr[O\rho],
  \qquad (\rho,O)\in X.
\]
This deterministic value is the deviation from the ideal
identity execution; finite-shot measurements yield a random
estimate $\widehat F$.
\end{definition}

\begin{definition}[Fingerprint Distance]\label{def:order}
For two fingerprints over the same interface $X$, their distance is
\[
  d_X(F_1,F_2)=\|F_1-F_2\|_\infty
  =\max_{x\in X}|F_1(x)-F_2(x)|.
\]
\end{definition}

An interface can be refined by adding experiments. If $X\subseteq X'$, the fingerprint over $X$ is the restriction of the fingerprint over $X'$ to the smaller interface. We write
\begin{equation}\label{eq:refine}
  F\sqsubseteq_{\mathcal{F}}F'
  \quad\text{iff}\quad
  X\subseteq X'\text{ and }F=F'|_X.
\end{equation}
Thus, refinement retains all previously recorded observations and adds new observable information.

\begin{definition}[Concrete and Abstract Domains]\label{def:concrete-abstract}
The concrete domain $(\Chan,\sqsubseteq_C)$ consists of the CPTP maps on $n$ qubits, equipped with an information order. The abstract domain $(A_X,\sqsubseteq_A)$ consists of fingerprints over $X$, also equipped with an information order. The abstraction map $\alpha_X:\Chan\to A_X$ sends a channel to its exact interface fingerprint, namely the expectation-value deviations that unlimited repetitions of the selected experiments would reveal.
\end{definition}

The concrete and abstract orders used in our probabilistic Galois-connection instance are specified in \cref{sec:concrete:pgc}. The exact fingerprint itself does not depend on those orders or on a particular measurement procedure.

\subsection{Probabilistic Concretisation}
\label{subsec:probabilistic-concretisation}

A fingerprint is an abstract description of a channel. To relate this description to concrete channels, we use a probabilistic concretisation: an abstract element may denote a distribution over concrete objects rather than a single object. This extends the classical Galois-connection formulation while retaining an order-theoretic relationship through the expectation of the distribution.

\begin{definition}[Galois Connection]\label{def:classical-galois}
Let $(C,\sqsubseteq_C)$ and $(A,\sqsubseteq_A)$ be partially ordered sets. Maps $\alpha:C\to A$ and $\gamma:A\to C$ form a Galois connection if, for every $c\in C$ and $a\in A$,
\[
  \alpha(c)\sqsubseteq_A a
  \quad\Longleftrightarrow\quad
  c\sqsubseteq_C\gamma(a).
\]
\end{definition}

Let $C$ be a subset of a real vector space, and let $\mathcal{D}(C)$ denote the set of probability distributions over $C$ whose expectation exists and belongs to $C$ (see \cref{sec:app:prob}).

\begin{definition}[Point Mass]\label{def:point-mass}
For any $c\in C$, the point mass at $c$, denoted by $[c]$, is the probability distribution over $C$ defined, for every measurable set $B\subseteq C$, by
\[
  [c](B)=
  \begin{cases}
    1 & \text{if }c\in B,\\
    0 & \text{otherwise.}
  \end{cases}
\]
In particular, if $Y\sim[c]$, then $\mathbb{E}[Y]=c$.
\end{definition}

\begin{definition}[Probabilistic Concretisation]\label{def:probabilistic-concretisation}
A probabilistic concretisation is a map $\gamma_P:A\to\mathcal{D}(C)$. Its expected concretisation is the map $\bar\gamma_P:A\to C$ defined by
\[
  \bar\gamma_P(a)=\mathbb{E}[Y],\qquad Y\sim\gamma_P(a).
\]
\end{definition}

The distribution $\gamma_P(a)$ captures uncertainty about which concrete object corresponds to an abstract element—for example, modelling a noisy channel as a random mixture of ideal operations. Its expected concretisation is a concrete element and can thus be compared using the concrete order.

\begin{definition}[Probabilistic Galois Connection]\label{def:galois}
The pair $(\alpha,\gamma_P)$ is an expectation-based probabilistic Galois connection if, for every $c\in C$ and $a\in A$,
\[
  \alpha(c)\sqsubseteq_A a
  \quad\Longleftrightarrow\quad
  c\sqsubseteq_C\bar\gamma_P(a).
\]
\end{definition}

\begin{theorem}[Generalisation of the classical connection]\label{thm:pgc-generalises}
\textup{(1)} If $(\alpha,\gamma_P)$ is a probabilistic Galois connection, then $(\alpha,\bar\gamma_P)$ is a classical Galois connection.
\textup{(2)} If $(\alpha,\gamma)$ is a classical Galois connection, then $(\alpha,\gamma_P)$ with $\gamma_P(a)=[\gamma(a)]$ is a probabilistic Galois connection.
\end{theorem}
\begin{proof}
For (1), the defining law of \cref{def:galois} is precisely the classical Galois law for $\alpha$ and $\bar\gamma_P$. For (2), the expectation of the point mass $[\gamma(a)]$ is $\gamma(a)$, so the probabilistic law reduces to the classical one.
\end{proof}

The classical definition is therefore recovered exactly when every concretisation is a point mass. The probabilistic law depends on $\gamma_P$ only through its expected value: distributions with the same expectation are indistinguishable under this order-theoretic relation. As in the classical case, instantiating the law with $a=\alpha(c)$ gives $c\sqsubseteq_C\bar\gamma_P(\alpha(c))$.

\subsection{Observational Soundness}
\label{subsec:soundness}

The semantic guarantee of a fingerprint is relative to its experiment interface. For two channels, define their observational discrepancy over $X$ by
\[
  \Delta_{\mathrm{obs}}(\Lambda_1,\Lambda_2)
  =\max_{(\rho,O)\in X}
  \left|\Tr[O\Lambda_1(\rho)]-\Tr[O\Lambda_2(\rho)]\right|.
\]
The ideal expectation in the fingerprint definition is common to both channels and therefore cancels when their fingerprints are compared.

\begin{theorem}[Exact observational soundness]\label{prop:exact-sound}
For any $\Lambda_1,\Lambda_2\in\Chan$,
\[
  \|\alpha_X(\Lambda_1)-\alpha_X(\Lambda_2)\|_\infty
  =\max_{(\rho,O)\in X}
  \left|\Tr[O\Lambda_1(\rho)]-\Tr[O\Lambda_2(\rho)]\right|
  =\Delta_{\mathrm{obs}}(\Lambda_1,\Lambda_2).
\]
\end{theorem}
\begin{proof}
For every $(\rho,O)\in X$, subtracting the two entries of the fingerprints cancels the common ideal term $\Tr[O\rho]$. Taking the maximum absolute value over $X$ gives the stated equality.
\end{proof}

Consequently, fingerprints at distance at most $\varepsilon$ represent channels whose outcomes differ by at most $\varepsilon$ on every experiment in $X$. Equal fingerprints imply observational equivalence on that interface. They do not imply equivalence under experiments outside it: distinguishing such channels requires refining the interface as in \cref{eq:refine}.

The observational distance is also bounded by the operational channel distance.

\begin{definition}[Lipschitz Continuity]\label{def:lipschitz}
The abstraction $\alpha_X$ is Lipschitz-continuous with constant $L$ if
\[
  \|\alpha_X(\Lambda_1)-\alpha_X(\Lambda_2)\|_\infty
  \leq L\,\|\Lambda_1-\Lambda_2\|_\diamond
\]
for all channels $\Lambda_1,\Lambda_2$, where $\|\cdot\|_\diamond$ is the diamond norm, the standard operational distance between channels~\cite{watrous_simpler_2013,watrous2018theory,Benenti_2010}.
\end{definition}

\begin{proposition}[Diamond-norm upper bound]\label{prop:diamond-upper}
For the normalised observables of \cref{def:reference}, the exact fingerprint abstraction is Lipschitz-continuous with $L\leq1$:
\[
  \|\alpha_X(\Lambda_1)-\alpha_X(\Lambda_2)\|_\infty
  \leq \|\Lambda_1-\Lambda_2\|_\diamond.
\]
\end{proposition}
\begin{proof}
Per input state $\rho$ and observable $O$ with $\|O\|_{\mathrm{op}}\leq1$, the expectation-value difference is bounded by the operational distinguishability of the two channel outputs, and hence by the diamond norm of $\Lambda_1-\Lambda_2$. Taking the maximum over the selected pairs $(\rho,O)\in X$ proves the claim.
\end{proof}
\looseness=-1

The bound is interface-relative: the diamond norm takes a supremum over all admissible inputs (including inputs entangled with an ancillary system), whereas the fingerprint takes a maximum only over the selected states and observables. The two distances can therefore coincide for some channel pairs and differ for others.

The abstraction must satisfy four complementary properties, stated in the remaining of this section, independently of any particular measurement protocol used to estimate a fingerprint.

\subsection{Conciseness}
\label{subsec:conciseness}

A general CPTP map acting on $n$ qubits requires $\Theta(16^n)$ real parameters. A fingerprint instead stores one scalar for each state--observable pair in $X$.

\begin{definition}[Conciseness]\label{def:conciseness}
The abstraction is concise if its representation size is negligible relative to a full channel representation. For the interface $X=\States\times\Obs$,
\[
  |\alpha_X(\Lambda)|=|\States|\cdot|\Obs|=|X|,
  \qquad \dim(\Chan_n)=\Theta(16^n),
\]
where $\Chan_n$ is the set of channels on $n$-qubit systems.
\end{definition}

Storage cost scales linearly with interface size, independent of the channel space’s intrinsic dimension. If the interface grows polynomially with the number of qubits, the fingerprint is exponentially smaller than a full channel representation. This intentional compression preserves only the interface-selected observations, not all possible channel behaviour. Refining the interface keeps existing entries and adds observable information, increasing representation size accordingly.

\subsection{Estimation Efficiency}
\label{subsec:efficiency}

A compact abstract representation is not sufficient if obtaining it requires an excessive number of measurements. We therefore state estimation efficiency as a requirement on the distribution of measured fingerprints, independently of any particular estimator.

\begin{definition}[Estimation Efficiency]\label{def:est-eff}
Let $\alpha_T(\Lambda)$ be the distribution of fingerprints of $\Lambda$ obtained using $T$ measurement rounds. The abstraction admits efficient estimation if, for every accuracy $\varepsilon>0$ and failure probability $\delta\in(0,1)$, a number of rounds $T$ polynomial in $|X|$, $1/\varepsilon$ and $\log(1/\delta)$, together with polynomial classical post-processing, ensures that for every channel $\Lambda$,
\[
  \Pr_{F\sim\alpha_T(\Lambda)}
  \left[\|F-\alpha_X(\Lambda)\|_\infty\leq\varepsilon\right]
  \geq1-\delta.
\]
\end{definition}
The tolerance $\varepsilon$ bounds the largest entrywise error, while $\delta$ bounds the probability that the complete fingerprint exceeds that tolerance. The probability is over measurement outcomes, not over channels. A probabilistic guarantee allows rare deviations to be accounted for through $\delta$ rather than forcing the tolerance to cover every possible measurement outcome. 
For a measured fingerprint \(F\), its uniform estimation error \(D_\Lambda(F)\) is given by
\[
  D_\Lambda(F)=\|\alpha_X(\Lambda)-F\|_\infty.
\]

\begin{proposition}[Tolerance at a given confidence]\label{prop:pgc-tolerance}
\textup{(1)} If every $\alpha_T(\Lambda)$ is a point mass, the guarantee of \cref{def:est-eff} does not depend on $\delta$.\
\textup{(2)} The guarantee holds for $(\varepsilon,\delta)$ if and only if, for every $\Lambda$, $\varepsilon$ is at least the $(1-\delta)$-quantile of $D_\Lambda$ under $\alpha_T(\Lambda)$.
\end{proposition}
\begin{proof}
For (1), under a point mass, the event has probability either zero or one, and $1-\delta>0$. For (2), the required condition is $\Pr[D_\Lambda\leq\varepsilon]\geq1-\delta$, which is equivalent to $\varepsilon$ being at least the corresponding quantile (see \cref{sec:app:prob}).
\end{proof}
For comparison, if a protocol estimates one entry by the mean of $T$ independent outcomes in $[a,b]$, Hoeffding's inequality gives $\delta=2e^{-2T\varepsilon^2/(b-a)^2}$. For outcomes in $[-1,1]$, $T=1000$ and $\varepsilon=0.1$ give $\delta\approx0.013$. A concrete construction must additionally state the computational and measurement resources needed to achieve its uniform guarantee; \cref{sec:concrete} provides such a construction.

\subsection{Adaptiveness and Stability}
\label{subsec:adaptiveness}

Noise can evolve over time, so useful fingerprints should change continuously with the underlying channel and should be updatable as additional observations arrive. We separate continuity of the exact abstraction from convergence of its estimator.

\begin{definition}[Continuity]\label{def:continuity}
The abstraction is continuous if small perturbations in $\Lambda$ induce small changes in $\alpha_X(\Lambda)$ under the uniform norm $\|\cdot\|_\infty$.
\end{definition}

Continuity ensures that gradual changes in device noise produce gradual changes in the exact fingerprint rather than abrupt discontinuities.

\begin{definition}[Incremental Estimator Convergence]\label{def:incremental}
Let $\hat\alpha_T$ be an empirical fingerprint estimator based on $T$ rounds. The estimator has incremental convergence if its expected uniform error satisfies:
\[
  \mathbb{E}\left[\|\hat\alpha_T-\alpha_X(\Lambda)\|_\infty\right]
  =O\!\left(\frac{1}{\sqrt{T}}\right).
\]
\end{definition}

This is the usual statistical convergence rate as additional samples are collected under a fixed channel; it does not guarantee that observations collected before a channel change remain unbiased for the new one. The effect of reusing stale observations is evaluated under controlled drift in \cref{sec:res:rq4}.

\begin{definition}[Robustness]\label{def:robustness}
The abstraction is robust if an estimation error of at most $\varepsilon$ implies that the corresponding observable deviations are bounded by $\varepsilon$:
\[
  \|\hat\alpha-\alpha_X(\Lambda)\|_\infty\leq\varepsilon
  \quad\Longrightarrow\quad
  \text{observable deviations are bounded by }\varepsilon.
\]
\end{definition}

By \cref{prop:exact-sound}, this robustness property is interface-relative: it applies to every experiment in $X$, and makes no claim about observations omitted from the interface.

\section{\tool: Fingerprint Construction via Classical Shadows}
\label{sec:concrete}
\label{sec:fingerprints}


This section instantiates the exact fingerprint abstraction $\alpha_X$ (defined in \cref{sec:semantics}) as $\alpha_{\mathrm{shadow}}$ using classical shadows~\cite{huang2020predicting} and the live-fingerprinting workflow in~\cite{simshadow:ICST:2026}. The construction has three components: a finite probe suite instantiates the experiment interface; shared randomised measurements estimate all observables jointly; and a median-of-means estimator provides a uniform finite-sample guarantee. Reusing a single measurement ensemble per reference state and for all observables amortises acquisition across the interface and enables fingerprint refreshing under changing noise. Channel reconstruction from a fingerprint is optional (See Appendix, \cref{sec:concrete:gamma}).


\subsection{Probe Suite: Instantiating the Interface}
\label{sec:concrete:probe}

An experiment interface $X=\States\times\Obs$ (\cref{def:reference}) is determined by its reference states, observables, and the locality of those observables. We use the maximum Pauli weight:
\[
  w^\star=\max_{O\in\Obs}w(O),
\]
where $w(O)$ is the number of non-identity factors in a Pauli string $O$. A \textit{Pauli String} is a tensor product with one operator per qubit, chosen from \(I\), \(X\), \(Y\), and \(Z\). For example, \(ZZI\) means \(Z\otimes Z\otimes I\).

\mypara{Reference states}
The probe suite used in this work is adapted from~\cite{simshadow:ICST:2026}. It combines three types of reference states, each exposing different aspects of noise:
\begin{itemize}[leftmargin=*,noitemsep,topsep=0pt]
  \item computational-basis states (\eg{} $\ket{0\cdots0}$ and $\ket{1\cdots1}$), which expose population errors;
  \item product superposition states (\eg{} $\ket{+}\ket{0}$ and $\ket{0}\ket{-}$), which expose coherence errors;
  \item entangled states (\eg{} $\ket{\Phi^+}$), which expose correlated multi-qubit errors.
\end{itemize}

\mypara{Observables and interface size}
The observable set consists of bounded-weight Pauli strings
\[
  O=\bigotimes_q P_q,\qquad P_q\in\{I,X,Y,Z\},
\]
including two-qubit correlators. For $k=|\States|$ reference states and $m=|\Obs|$ observables, the interface is $X=\States\times\Obs$, thus $|X|=k\cdot m$.

The values of $k$, $m$, and $w^\star$ are design parameters. They can be chosen to match the diagnostic objective while keeping the interface finite. The experiments below use the relevant finite instantiations or subsets of this probe-suite design, as specified in each experimental setup.

\subsection{Shared Classical-Shadow Measurements}
\label{sec:concrete:shadow}

Fix a reference state $\rho_i=\ket{\psi_i}\!\bra{\psi_i}\in\States$. Applying the channel under test $\Lambda\in\Chan$ produces the state $\Lambda(\rho_i)$. We collect $T$ independent measurement rounds for each reference state. In every round, all $n$ qubits are measured in parallel in independently sampled single-qubit Pauli bases.

For round $t$, sample a product basis%
\[
  U^{(t)}=\bigotimes_{q=1}^{n}U_q^{(t)},
\]
where each $U_q^{(t)}$ is drawn uniformly from $\{H,HS^\dagger,I\}$, rotating the $X$, $Y$, or $Z$ eigenbasis onto the computational basis. Measure $U^{(t)}\Lambda(\rho_i)U^{(t)\dagger}$ in the computational basis, obtaining a bitstring $b^{(t)}\in\{0,1\}^n$ and its associated eigenvalue signs $s^{(t)}\in\{+1,-1\}^n$. The corresponding classical-shadow snapshot is
\[
  \widehat\rho_i^{(t)}
  =\bigotimes_{q=1}^{n}
  \left(3U_q^{(t)\dagger}\ket{b_q^{(t)}}\!\bra{b_q^{(t)}}U_q^{(t)}-I\right),
\]
which is the inverse of the single-qubit measurement channel applied independently on each qubit~\cite{huang2020predicting}.
The snapshot is an unbiased estimator of the post-channel state:
\[
  \mathbb{E}[\widehat\rho_i^{(t)}]=\Lambda(\rho_i).
\]
Consequently, for every observable $O_j$, the single-round score
\[
  \widehat o_{ij}^{(t)}=\Tr\!\left(O_j\widehat\rho_i^{(t)}\right)
\]
is unbiased for $\Tr[O_j\Lambda(\rho_i)]$. For a Pauli string $O_j=\bigotimes_qP_q$, the score has the closed form
\begin{equation}\label{eq:single-shot}
  \widehat o_{ij}^{(t)}=
  \begin{cases}
    \displaystyle\prod_{q:P_q\neq I}3s_q^{(t)},
      &\text{if the sampled basis matches $P_q$ for every $q$ with $P_q\neq I$,}\\[2pt]
    0,&\text{otherwise.}
  \end{cases}
\end{equation}
Its single-round variance is the shadow norm of $O_j$:
\begin{equation}\label{eq:shadow-norm}
  \Var(\widehat o_{ij}^{(t)})
  =3^{w(O_j)}-\Tr\!\left(O_j\Lambda(\rho_i)\right)^2.
\end{equation}

Crucially, the same $T$ snapshots $\{\widehat\rho_i^{(t)}\}_{t=1}^{T}$ are reused to estimate every observable $O_j\in\Obs$. Measurement is therefore amortised across the $m$ observables rather than repeated separately for each one. Since the scores in \cref{eq:single-shot} can have heavy tails for higher-weight observables, we aggregate them using the median-of-means estimator~\cite{huang2020predicting}. 

We partition the $T$ rounds into $K$ equal batches, average the scores within each batch, and take the median of the batch means $\widehat\mu_{ij}$, for statistical robustness \cite{simshadow:ICST:2026}, thus
\begin{equation}\label{eq:mom}
  \widehat\mu_{ij}
  =\operatorname{median}_{1\leq b\leq K}
  \left(\frac{K}{T}\sum_{t\in\text{batch }b}\widehat o_{ij}^{(t)}\right), \quad   K=\left\lceil2\log\!\left(\frac{2|X|}{\delta}\right)\right\rceil
\end{equation}

\subsection{Deviation Fingerprint and Estimation Guarantee}
\label{sec:concrete:alpha}
Each entry records the deviation of the estimated output expectation from its ideal, noiseless value. For the prepared pure state $\rho_i=\ket{\psi_i}\!\bra{\psi_i}$, the empirical fingerprint entry is
\begin{equation}\label{eq:fingerprint-entry}
  \widehat F_{ij}=\widehat\mu_{ij}-\mu_{ij}^{\mathrm{ideal}},
  \qquad \text{where}\quad
  \mu_{ij}^{\mathrm{ideal}}=\bra{\psi_i}O_j\ket{\psi_i}.
\end{equation}
Collecting these entries gives $\widehat F\in\mathbb{R}^{X}$, an estimate of the exact fingerprint $\alpha_X(\Lambda)$ defined in \cref{def:exact-fingerprint}. The construction is denoted by $\alpha_{\mathrm{shadow}}$; its output is random because it depends on the sampled measurement bases and outcomes.

While the single-round score is unbiased, it is generally not the case for the median of means. For example, for $\Lambda(\rho)=\ket0\!\bra0$ and $O=Z$, a single round returns $3$ with probability $1/3$ and $0$ otherwise, giving the median of three one-round batches expectation $7/9$ rather than $1$. All guarantees below rely on the estimator's concentration bound, not unbiasedness.

\begin{algorithm}[t]
\caption{$\alpha_{\mathrm{shadow}}$: shared classical-shadow fingerprint of a channel $\Lambda$}
\label{alg:alpha-shadow}
\label{alg:alpha-shadow-full}
\footnotesize
\begin{algorithmic}[1]
\Require Reference states $\States=\{\rho_i\}_{i=1}^{k}$, Pauli observables $\Obs=\{O_j\}_{j=1}^{m}$, rounds $T$ per reference state, and number of batches $K$
\Ensure Estimated fingerprint $\widehat F\in\mathbb{R}^{k\times m}$
\For{each reference state $\rho_i\in\States$}
  \State Prepare $\rho_i$ and apply $\Lambda$
  \For{$t=1$ to $T$}
    \Comment{One shared round measures all qubits in parallel}
    \State Sample a product Pauli basis $U^{(t)}=\bigotimes_qU_q^{(t)}$ independently on each qubit
    \State Measure $U^{(t)}\Lambda(\rho_i)U^{(t)\dagger}$ in the computational basis, obtaining $s^{(t)}\in\{\pm1\}^{n}$
  \EndFor
  \For{each observable $O_j=\bigotimes_qP_q\in\Obs$}
    \Comment{Reuse the same $T$ rounds for every observable}
    \State Compute the scores $\widehat o_{ij}^{(t)}$ using \cref{eq:single-shot}
    \State $\widehat\mu_{ij}\gets\operatorname{MedianOfMeans}(\{\widehat o_{ij}^{(t)}\}_{t=1}^{T},K)$ using \cref{eq:mom}
    \State $\widehat F_{ij}\gets\widehat\mu_{ij}-\bra{\psi_i}O_j\ket{\psi_i}$ using \cref{eq:fingerprint-entry}
  \EndFor
\EndFor
\State \Return $\widehat F$
\end{algorithmic}
\end{algorithm}

The exact semantic guarantee for this abstraction was established in \cref{prop:exact-sound}: the exact fingerprint distance equals the largest observational discrepancy over $X$. The role of the shadow construction is to estimate that exact fingerprint with a finite measurement budget.

\begin{proposition}[Uniform estimation guarantee of $\alpha_{\mathrm{shadow}}$]\label{prop:alpha-shadow}
Fix an accuracy $\varepsilon>0$, confidence level $1-\delta$, and a probe suite whose observables have maximum Pauli weight $w^\star$. With
\[
  T=O\!\left(3^{w^\star}\varepsilon^{-2}
  \log\!\left(\frac{|X|}{\delta}\right)\right)
\]
classical-shadow rounds per reference state and the median-of-means estimator of \cref{eq:mom}, the empirical fingerprint satisfies the following, with probability at least $1-\delta$:
\[
  \|\alpha_{\mathrm{shadow}}(\Lambda)-\alpha_X(\Lambda)\|_\infty
  \leq\varepsilon.
\]
\end{proposition}

The factor $3^{w^\star}$ captures the dependence on the maximum Pauli weight, with $w^\star$ fixed by design in our probe suites. The following corollary gives the explicit tolerance used in the evaluation.

\begin{corollary}[Estimation guarantee of \tool{}]\label{cor:pgc-shadow}
Let $\alpha_T(\Lambda)$ be the distribution of $\alpha_{\mathrm{shadow}}(\Lambda)$ computed from $T$ rounds per reference state with $K$ median-of-means batches. Then the guarantee of \cref{def:est-eff} holds for
{
     \setlength{\abovedisplayskip}{0pt}
     \setlength{\belowdisplayskip}{1pt}
     \setlength{\abovedisplayshortskip}{0pt}
     \setlength{\belowdisplayshortskip}{1pt}
     \small
\[
  \varepsilon(T,\delta)
  =\sqrt{\frac{34\cdot3^{w^\star}}{\lfloor T/K\rfloor}}
  \quad \texttt{where} \quad   K=\left\lceil2\log\!\left(\frac{2|X|}{\delta}\right)\right\rceil.
\]\normalsize}%
\end{corollary}
\begin{proof}
With batches of at least $\lfloor T/K\rfloor\geq34\cdot3^{w^\star}/\varepsilon^2$ rounds, the median-of-means guarantee of~\cite{huang2020predicting} (Theorem~1; the shadow norm of a weight-$w$ Pauli observable is $3^w$) gives uniform error at most $\varepsilon$ with probability at least $1-\delta$. The choice of $K$ allocates failure probability at most $\delta/k$ to the $m$ observables for each of the $k$ reference states; a union bound over the states gives total failure probability at most $\delta$. This is the condition of \cref{def:est-eff}.
\end{proof}

The certified tolerance is conservative. By \cref{prop:pgc-tolerance}, the smallest valid tolerance is the $(1-\delta)$-quantile of the uniform estimation error, which we measure in \cref{sec:res:rq1}.

\subsection{Information Orders and the Pauli-Channel Concretisation}
\label{sec:concrete:pgc}

We instantiate the information orders of \cref{def:concrete-abstract} using the orders of Coecke and Martin~\cite{Coecke2011}. Under these orders, the completely mixed state is least informative and pure states are maximal.

\mypara{Orders on states}
For probability vectors, the Bayesian order is defined by $x\sqsubseteq y$ if some permutation makes both $x$ and $y$ non-increasing and $x_iy_{i+1}\leq x_{i+1}y_i$ for every $i$. The uniform distribution is its least element and point masses are maximal. For two outcomes, this reduces to
\[
  x\sqsubseteq y
  \quad\text{iff}\quad
  y_1\leq x_1\leq\tfrac12
  \quad\text{or}\quad
  \tfrac12\leq x_1\leq y_1.
\]
For density matrices, the spectral order $r\sqsubseteq s$ requires that $r$ and $s$ admit a common eigenbasis in which their eigenvalue lists satisfy the Bayesian order. Requiring commutation is necessary for this to define an information order~\cite{Coecke2011}.

\mypara{Orders on channels and fingerprints}
Channels are ordered by their outputs on the reference states:
\[
  \Lambda_1\sqsubseteq_C\Lambda_2
  \quad\text{iff}\quad
  \Lambda_1(\rho)\sqsubseteq\Lambda_2(\rho)
  \text{ in the spectral order for every }\rho\in\States.
\]
Channels with equal outputs on every probe are identified. Each experiment $(\rho,O)$ with a Pauli observable has two outcomes, so a fingerprint entry determines a two-outcome distribution. Fingerprints are ordered entrywise by the Bayesian order on these distributions. Writing $E=\Tr[O\Lambda(\rho)]$ for the expectation value behind an entry, this order is
\[
  E_1\sqsubseteq E_2
  \quad\text{iff}\quad
  E_1E_2\geq0\text{ and }|E_1|\leq|E_2|,
\]
so noise moves expectation values towards zero without changing their sign.

\mypara{Probabilistic concretisation for Pauli channels}
A single-qubit Pauli channel
\[
  \Lambda(\rho)
  =(1-p_x-p_y-p_z)\rho
  +p_xX\rho X+p_yY\rho Y+p_zZ\rho Z
\]
is a random mixture of unitary channels. We take that mixture as its probabilistic concretisation:
\[
  \gamma_P(a)
  =(1-p_x-p_y-p_z)[\mathcal I]
  +p_x[\mathcal X]+p_y[\mathcal Y]+p_z[\mathcal Z],
\]
where $[\mathcal U]$ is the point mass at the unitary channel $\rho\mapsto U\rho U^\dagger$. The rates are determined by the fingerprint. In particular, if $q_P$ is the probability that the $+1$ eigenstate of $P$ is found flipped in its own basis, then
$q_Z=p_x+p_y$, $q_X=p_y+p_z$ and $q_Y=p_x+p_z$,
which determine $(p_x,p_y,p_z)$. The expected concretisation $\bar\gamma_P(a)$ is the Pauli channel with these rates.

\begin{proposition}[Galois law for Pauli channels]\label{prop:pgc-pauli}
On the single-qubit interface of \cref{sec:concrete:probe}, with $\Chan$ the set of Pauli channels and $A=\alpha_X(\Chan)$, the pair $(\alpha_X,\gamma_P)$ is a probabilistic Galois connection, and for every Pauli channel $\Lambda$ we have
\[  \bar\gamma_P(\alpha_X(\Lambda))=\Lambda.\]
\end{proposition}
\begin{proof}
A Pauli channel maps each Pauli eigenstate $\rho$ to a state diagonal in the eigenbasis of $\rho$, with spectrum $(1-q,q)$, where $q$ is the total rate of errors that anticommute with the corresponding Pauli operator. Outputs of two Pauli channels on $\rho$ therefore commute, and their spectral order is the Bayesian order of these spectra, which are the outcome distributions of measuring $\rho$ in its own basis. Measuring either of the other two Pauli observables gives the uniform distribution for every Pauli channel. Hence $\alpha_X(\Lambda_1)\sqsubseteq_A\alpha_X(\Lambda_2)$ iff $\Lambda_1\sqsubseteq_C\Lambda_2$. The flip probabilities determine the rates, so $\bar\gamma_P(\alpha_X(\Lambda))=\Lambda$ and $\alpha_X(\bar\gamma_P(a))=a$ for every $a\in A$. The probabilistic Galois law follows.
\end{proof}

For channels that are not Pauli channels, including amplitude damping and the hardware-calibrated models, outputs on some probes do not commute. Distinct channels are then incomparable in the spectral order, and the law holds only trivially. \Cref{sec:res:rq1} evaluates both cases.

\subsection{Measurement and Post-Processing Cost}
\label{sec:concrete:cost}
Each reference state requires its own set of measurement rounds. By \cref{prop:alpha-shadow}, the total number of measurements is
\[
  O\!\left(
    |\States|\cdot3^{w^\star}\cdot\varepsilon^{-2}
    \log\!\left(\frac{|X|}{\delta}\right)
  \right).
\]


Classical post-processing to construct the fingerprint requires $O(T|X|)$ operations, with the measurement budget scaling in the number of reference states while records for each state are shared across all observables in the interface. The direct workflow stores only $|X|$ scalar deviations, avoiding the construction of a full $2^n\times2^n$ density or $4^n\times4^n$ Choi matrix. Channel reconstruction via Choi-matrix optimisation is not required for fingerprint comparison, but is described in \cref{sec:concrete:gamma}.

\section{Evaluation}
\label{sec:eval}

We evaluate the concrete construction $\alpha_{\mathrm{shadow}}$ against the abstraction properties defined in \cref{sec:properties}. The evaluation combines exact calculations from channel representations with empirical estimates obtained from randomized measurements. 
It is organised around four questions, each targeting a different aspect of the abstraction, with \autoref{tab:rq-mapping} mapping each question to its formal property, metric, and theoretical target:
\begin{itemize}[leftmargin=*,noitemsep,topsep=0pt]
  \item \textbf{RQ1 (Observational fidelity).} How accurately does fingerprint distance represent the observable discrepancy between concrete channels, and how reliable are finite-shot fingerprints?
  \item \textbf{RQ2 (Conciseness).} How does the representation size compare with a full channel representation, and how much information about the tested noise parameters is retained?
  \item \textbf{RQ3 (Measurement efficiency).} How does the number of shared measurement rounds required for a uniform error target grow with the interface size?
  \item \textbf{RQ4 (Adaptation under drift).} When noise changes, how does estimation converge, and when does reusing previous snapshots reduce the fresh measurement cost?
\end{itemize}
\subsection{Experimental Methodology}
\label{sec:eval:method}
\label{sec:methodology}

\subsubsection{Analytic and Empirical Tracks}
\label{sec:method:design}

We use a dual-track design. The \emph{analytic track} computes exact fingerprints directly from the Kraus representation of the underlying quantum channel (eq. \ref{eq:kraus}). For RQ1 and RQ4, channel distances are additionally computed using the diamond norm $\|\Lambda_1-\Lambda_2\|_\diamond$, evaluated by the Watrous \textit{Semi-Definite Program} (SDP)~\cite{watrous_simpler_2013}.

The \emph{empirical track} runs the 
classical-shadow estimator $\alpha_{\mathrm{shadow}}$ of \cref{alg:alpha-shadow}, with parallel single-qubit Pauli measurements and median-of-means aggregation. For RQ1, RQ3, and RQ4, empirical fingerprints are compared against Kraus-derived fingerprints; 
RQ2 uses Kraus-derived fingerprints directly.
Where reported, the information-retention ratio is estimated with the Kraskov--St\"ogbauer--Grassberger $k$-nearest-neighbour mutual-information estimator~\cite{kraskov_estimating_2004}, and optional channel reconstruction uses the constrained SDP defined in \cref{sec:concrete:gamma}, eq. \ref{eq:gamma-sdp}. 

\begin{table*}[t!]
    \centering
    \footnotesize
  \caption{RQs: the formal property each targets, the metric used, and the
  theoretical target from \cref{sec:properties}.}
    \label{tab:rq-mapping}
  \begin{tabular}{llll}
    \toprule
    \textbf{RQ}  & \textbf{Property (Def.) }                        & \textbf{Metric}                                                        & \textbf{Theoretical target} \\
    \midrule
    RQ1 & Soundness (Def.~\ref{def:galois}, Thm.~\ref{prop:exact-sound}, Def.~\ref{def:lipschitz}) & law; coverage; $\rho(D_{\mathrm{abs}}, D_{\diamond})$; $L$ & law holds; cov.\ $\ge1-\delta$; $L\le 1$, high $\rho$ \\
    RQ2 & Conciseness (\cref{def:conciseness})    & $\lvert X\rvert$ vs $16^n$; $R_{\mathrm{info}}$               & exp.\ compression, $R_{\mathrm{info}}\!\approx\!1$ \\
    RQ3 & Efficiency (\cref{def:est-eff})         & $T$ vs $\lvert X\rvert$                                       & $O(\varepsilon^{-2}\log\lvert X\rvert)$ \\
    RQ4 & Adaptiveness (\cref{def:continuity,def:incremental,def:robustness}) & $\lVert \widehat{\Lambda}_T - \Lambda_T\rVert_\diamond$, slope & $O(T^{-1/2})$, $\Delta_T \ll T_{\mathrm{cold}}$ 
    \\ \bottomrule
  \end{tabular}
  \parskip=2pt 
\end{table*}

\subsubsection{Reference Implementation and Testbed}
\label{sec:testbed}

We simulate states and channels as double-precision density matrices, and represent quantum channels by their Kraus decompositions (eq. \ref{eq:kraus}). Unless stated otherwise, the baseline is the identity channel and the noise channels are depolarising, amplitude damping, and phase damping, assembled at the qubit level. 
The exact diamond norm is obtained from the Watrous SDP~\cite{watrous_simpler_2013} using an interior-point convex solver. The information-retention ratio uses the Kraskov--St\"ogbauer--Grassberger estimator~\cite{kraskov_estimating_2004}.

\mypara{Experiment interface}
The interface $X=\States\times\Obs$ follows the probe-suite design of \cref{sec:concrete:probe}. The single-qubit experiments use computational-basis and product-superposition reference states together with the Pauli observables $X$, $Y$, and $Z$. Multi-qubit experiments additionally use entangled reference states and bounded-weight Pauli strings, with $\|O\|_{\mathrm{op}}\leq1$. Unless an experiment states otherwise, the maximum observable weight is $w^\star=2$.

\mypara{Replication protocol}
Experiments involving empirical fingerprint estimation (RQ1, RQ3, and RQ4) are repeated independently \textbf{50} times to quantify sampling variability. Empirical results are reported as mean $\pm$ standard deviation where applicable. Shot budgets apply only to experiments involving empirical estimation; each question-specific budget is stated below.

\mypara{Artifact and execution environment}
All experiments are implemented in a self-contained reference package released as the accompanying \href{https://zenodo.org/records/23414157}{anonymised artifact}, with fixed dependency versions and random seeds, a containerised environment, and an automated test suite for reproducibility. Experiments were executed on an Apple MacBook Pro with an M1 Pro processor and 32~GB RAM.

\subsubsection{Question-Specific Setups}
\label{sec:exprqs}

Unless stated otherwise, the program $P$ is a single Hadamard gate. For parameterised noise channels, $p$ denotes the channel-strength parameter (\eg{} the depolarising probability), with the range specified by the experiment.

\mypara{RQ1: Observational fidelity}
We first evaluate soundness in a controlled one-qubit setting. The baseline is the noiseless channel, and the noise channel is single-qubit depolarising noise. We sweep the depolarising rate $p$ over $30$ equally spaced values in $[0,0.40]$. For each value, we compute the exact fingerprint and estimate an empirical fingerprint from $1500$ measurement shots. The empirical fingerprint distance is summarised by its mean and standard deviation and compared with the analytical diamond distance between the noiseless and depolarising channels.

\mypara{RQ2: Conciseness}
We evaluate conciseness through representation scaling and information retention. For scaling, we compare fingerprint-interface growth with the size of the concrete channel representation for $n\in\{1,\ldots,7\}$ qubits. For information retention, we compute fingerprints directly from the Kraus representations of the parameterised noise channels over $10$ equally spaced parameter values in $[0.05,0.35]$.

\mypara{RQ3: Measurement efficiency}
We use depolarising noise with strength $p=0.15$. The interface size $|X|$ is varied by increasing the number of reference states while keeping the observable set fixed to the three single-qubit Pauli observables. The shot budget is varied over 
$(50,100,200,350,\ldots,3500)$ to measure how the required number of rounds changes as the interface grows.

\mypara{RQ4: Adaptation under drift}
We simulate a time-varying depolarising channel whose noise strength $p(t)$ increases linearly with time. The interface is fixed at $|X|=18$, comprising $6$ reference states and $3$ single-qubit Pauli observables. We compare full fingerprint recomputation with incremental updates. For an update, the $T_{\mathrm{old}}$ snapshots collected before drift (the fingerprint at $t=0$) are pooled with $T_{\mathrm{new}}$ fresh snapshots from the post-drift channel using the count-weighted estimator implemented in the evaluation code. Because old snapshots come from the pre-drift channel, their contribution introduces bias relative to the post-drift fingerprint. We report the smallest $T_{\mathrm{new}}$ that restores the target accuracy.

\subsubsection{Expert-Signature Verification Protocol}
\label{sec:method:expert}

Expert-signature verification complements RQ1 by testing whether fingerprints reproduce the expected physical signatures of different noise processes, rather than just preserving channel distances (as measured in RQ1).

Numeric compliance with the abstraction bounds does not establish that a fingerprint reflects physically expected noise behaviour. We therefore apply a falsifiable, estimator-independent check in the style of metamorphic testing~\cite{pontolillo2025}: for each noise channel, domain knowledge predicts which state--observable pairs should show the largest deviations and which should remain invariant. 
The \href{https://zenodo.org/records/23414157}{oracle} directly encodes the expected physical behaviour of each noise channel.


\begin{definition}[Expert signature and verification]\label{def:expert-sig}
For a noise channel $M$, an \emph{expert signature} is a pair $(\mathcal{L}_M,\mathcal{Z}_M)$ of disjoint subsets of $X$: the salient pairs $\mathcal{L}_M$ should show a large deviation, while the invariant pairs $\mathcal{Z}_M$ should show none. Writing $F=\alpha_X(\Lambda_M)$, the concrete domain passes verification for $M$ at thresholds $(\tau,\tau_0)$ if
\[
  \min_{(\rho,O)\in\mathcal{L}_M}|F_{\rho O}|
  \geq \tau\max_{(\rho,O)\in X}|F_{\rho O}|
  \quad\text{and}\quad
  \max_{(\rho,O)\in\mathcal{Z}_M}|F_{\rho O}|\leq\tau_0.
\]
\end{definition}

We instantiate the two-qubit interface with states 
$ \{\ket{00},\ket{11},\ket{10},\ket{01},\ket{++},
  \ket{+1},\ket{0-},\ket{\Phi^+}\}$ and Pauli observables $\{ZZ,XX,YY,ZI,IZ,IX,XI,ZX\}$. We write $\ket{q_0q_1}$ for a computational-basis state and use the convention that an observable $AB$ acts as $A$ on qubit $0$ and $B$ on qubit $1$. Expert signatures follow from the standard physical arguments~\cite{nielsen2010quantum}: depolarising noise contracts Pauli expectations, so it is salient when the state is a stabiliser of the observable ($|\langle O\rangle|=1$) and invariant when $\langle O\rangle=0$; amplitude damping relaxes $\ket1$ to $\ket0$, so its population signatures appear on $Z$-type observables of the excited qubit; and phase damping erases coherences, so it affects $X/Y$-type observables of superposition and entangled states. We use $\tau=0.5$, $\tau_0=10^{-6}$, and channels at rate $0.1$. Fingerprints are computed exactly, so the verdict is independent of measurement noise.

\subsection{Results}
\label{sec:res}
\label{sec:results}
\label{sec:eval:results}

The exact-distance results (RQ1) establish what the interface preserves; 
RQ2-RQ4 measure the cost and practical behaviour of constructing and updating the corresponding fingerprints.

\begin{figure}[t]
    \begin{minipage}[t]{0.56\linewidth}
    \centering
    \captionsetup{hypcap=false}
    \includegraphics[width=0.64\linewidth] 
    {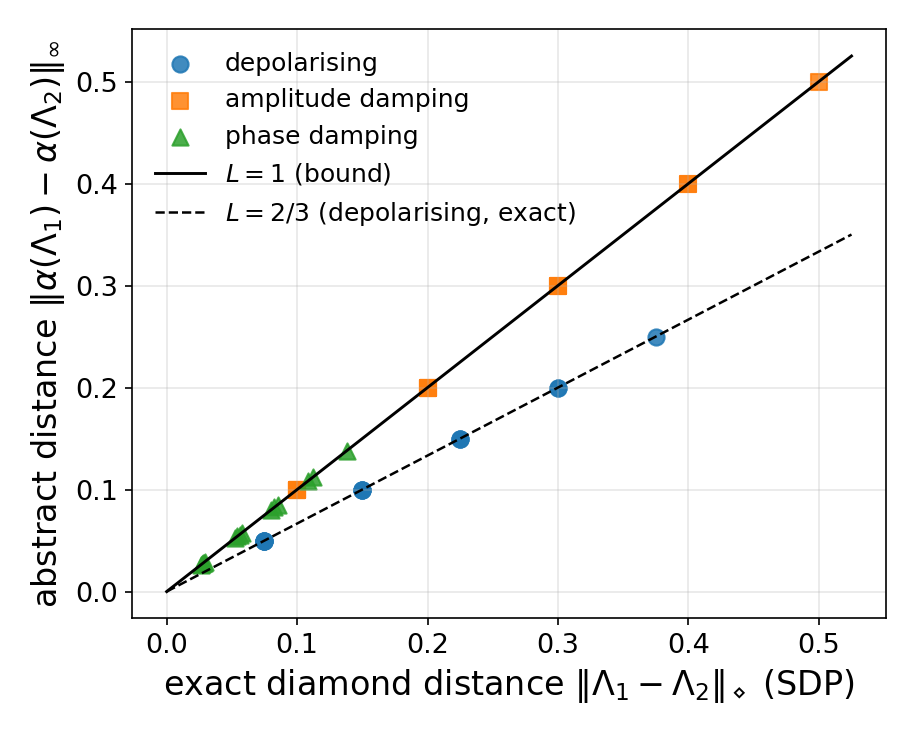}
    \Description{Scatter plot of abstract fingerprint distance against exact diamond distance for all channel pairs in three noise families; amplitude- and phase-damping pairs lie exactly on the identity line and depolarising pairs exactly on the two-thirds line.}
    \captionof{figure}{\textbf{RQ1 Abstract vs Concrete:} abstract distance vs SDP diamond distance for all within-family channel pairs. 
    Amplitude- and phase-damping pairs attain the soundness bound ($L=1$, solid); depolarising
    pairs lie exactly on $L=2/3$ (dashed), as the depolarising worst case is off the interface.}
    \label{fig:rq1}
    \end{minipage}
    \hfill
\begin{minipage}[t]{0.4\linewidth}
\captionsetup[figure]{hypcap=false}
\centering
\includegraphics[width=0.9\linewidth]
{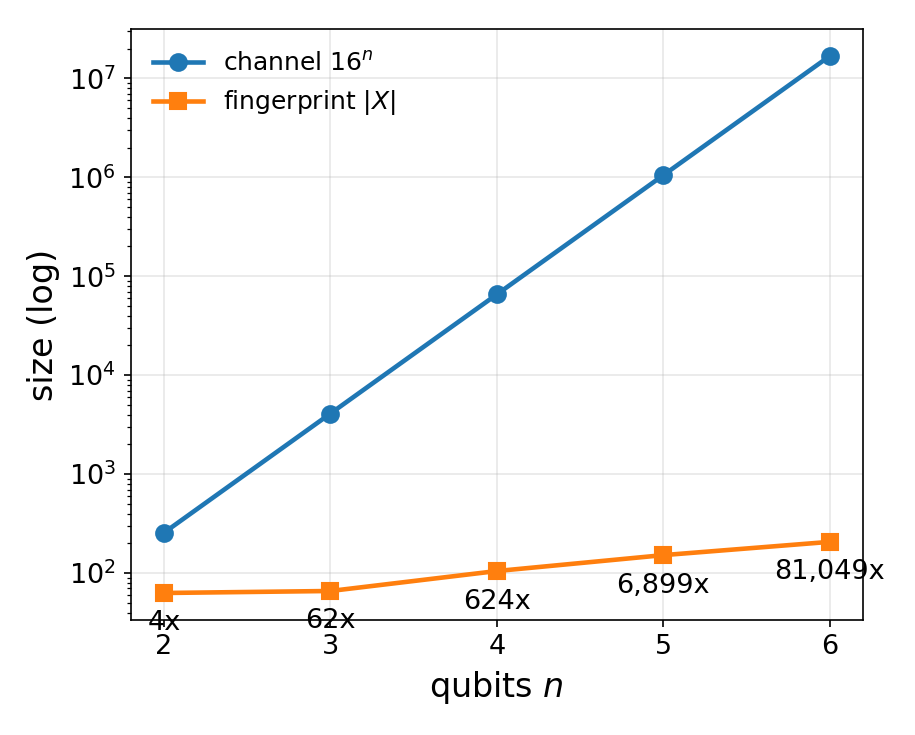}
\Description{Log-scale plot comparing the full channel
dimension $16^n$ and fingerprint-interface size
$\lvert X\rvert$ for $n=2$--$6$ qubits. Annotations
show the channel-to-fingerprint compression ratio
at each qubit count.}

\captionsetup{hypcap=false}
\captionof{figure}{\textbf{RQ2 Representation Size.}
Comparison of the full channel dimension ($16^n$)
and fingerprint-interface size ($\lvert X\rvert$)
for $n=2$--$6$ qubits on a logarithmic scale.
Annotations report the compression ratio
$16^n/\lvert X\rvert$.}
\label{fig:rq2}
\end{minipage}
\end{figure}
\looseness=-1

\subsubsection{RQ1: Observational Fidelity}
\label{sec:res:rq1}

\autoref{fig:rq1} compares abstract and concrete distances for all within-family pairs generated from
$p\in\{0.05,0.10,0.15,0.20,0.25,0.30\}$,
where $p$ is the strength parameter for depolarising, amplitude-damping, and phase-damping channels. The abstract and concrete distances correlate at $\rho=0.98$, and the realised Lipschitz constant is $L=1.00$. This matches the upper bound of \cref{prop:diamond-upper}.

The bound is attained for the amplitude- and phase-damping families: for these channels, the selected interface contains the diamond-norm worst-case direction and $D_{\mathrm{abs}}=D_{\diamond}$. For depolarising noise, the ratio is exactly $2/3$, since $D_{\mathrm{abs}}=|\Delta p|$ while $D_{\diamond}=\tfrac32|\Delta p|$; the diamond-norm worst case lies outside the selected interface. The restricted-interface experiment illustrates the same limitation directly: two channels can have abstract distance $0$ on $X'$ while their diamond distance is $0.163$. Thus, soundness is exact relative to the interface, not a claim of global channel equivalence.

\mypara{Probabilistic Galois connection}
On the single-qubit interface, we verify the law of \cref{def:galois} using the information orders and random-unitary concretisation of \cref{sec:concrete:pgc}. Over \GCpauliPairs{} pairs of \GCpauliChannels{} Pauli channels (random rates, depolarising, and dephasing families), the law holds in every case, with \GCpauliViolations{} violations. \GCpauliOrdered{} pairs are non-trivially ordered,
and the expected concretisation reconstructs every channel exactly. For \GCnonPauliChannels{} other channels (amplitude damping, IBM-calibrated, and random), abstract and concrete orders agree on all \GCnonPauliPairs{} pairs, but only trivially: no two distinct channels are ordered (\GCnonPauliDistinctOrdered{} ordered pairs), as their outputs on some probes do not commute.

\mypara{Finite-shot coverage}
We test the $(\varepsilon,\delta)$ guarantee of \cref{cor:pgc-shadow} on two interfaces: $|X|=18$ with $w^\star=1$, and the $117$-entry two-qubit suite with $w^\star=2$. The test uses four channels (depolarising, amplitude damping, phase damping, and IBM-calibrated), $\delta\in\{0.05,0.01\}$, $T\in\{2000,8000,32000\}$, and \PGCconfigs{} configurations with \PGCsessionsOne{} (one-qubit) or \PGCsessionsTwo{} (two-qubit) independent sessions each.

\emph{The certified tolerance is valid.} The measured fingerprint lies within the certified tolerance in every session (minimum coverage \PGCcertMinCov), and the Clopper--Pearson lower bound exceeds $1-\delta$ in every configuration. The certified tolerances are conservative: they range from \PGCepsCertMin{} to \PGCepsCertMax{} (and from \PGCepsCertMinTmax{} to \PGCepsCertMaxTmax{} at $T=32000$), exceeding the empirical $(1-\delta)$-quantile---the smallest valid tolerance by \cref{prop:pgc-tolerance}---by a factor of \PGCslackMin{}--\PGCslackMax{}.

\begin{table}[t]
  \centering
  \captionsetup{hypcap=false}
  \begin{minipage}[t]{0.47\linewidth}
    \vspace{0pt}
    \centering

  \centering
  \caption{RQ2: interface size $|X|$ compared with the full channel dimension $16^n$.\\}
  \label{tab:conciseness}
  \footnotesize
  \begin{tabular}{lrrr}
    \toprule
    $n$ & $|X|$ & $16^n$ & \textbf{Compression}\\
    \midrule
    2 & 63  & 256              & $4\times$\\
    3 & 66  & 4{,}096          & $62\times$\\
    4 & 105 & 65{,}536         & $624\times$\\
    5 & 152 & 1{,}048{,}576    & $6{,}899\times$\\
    6 & 207 & 16{,}777{,}216   & $81{,}049\times$\\
    \bottomrule
  \end{tabular}
\end{minipage}%
\hfill%
    \begin{minipage}[t]{0.5\linewidth}
    \centering
  \captionof{table}{RQ3: Single-round variance of classical-shadow estimator~(\cref{eq:single-shot}) with experiment pairs $(\rho,O)$ (Col. S,O) vs shadow norm $3^{w}-\langle O\rangle^2$~(\cref{eq:shadow-norm}) (Col. Theory); Col. Empirical is at $T=4{\times}10^4$. $w$ is as in \cref{sec:concrete:probe}.}
  \label{tab:variance}
  \footnotesize
  \begin{tabular}{lcrr}
    \toprule
    \textbf{$S,\Obs$} & \textbf{$w$} & \textbf{Theory} & \textbf{Empirical} \\
    \midrule
    $\ket{+}$, $Z$ & 1 & 3.00 & 3.01 \\
    $\ket{+}$, $X$ & 1 & 2.00 & 2.00 \\
    $\ket{\Phi^+}$, $ZZ$ & 2 & 8.00 & 7.98 \\
    $\ket{\Phi^+}$, $XX$ & 2 & 8.00 & 8.03 \\
    \bottomrule
  \end{tabular}
\Description{Table comparing theoretical and empirical single-round variances.}
  \end{minipage}%
\end{table}
\looseness=-1

\emph{The guarantee is informative at a tight tolerance.} Calibrating $\varepsilon$ as the $(1-\delta)$-quantile on half of the sessions and evaluating coverage on the other half gives mean held-out coverage \PGCholdoutFive{} for $\delta=0.05$ and \PGCholdoutOne{} for $\delta=0.01$, close to the nominal $1-\delta$.

\emph{The check has power.} A per-entry Gaussian interval ignoring the union over the interface violates the condition in \PGCentryFail{} of \PGCentryTotal{} two-qubit configurations (coverage \PGCentryMinCov{}--\PGCentryMaxCov{}), confirming that the guarantee concerns the complete fingerprint rather than its individual entries.  

Finally, over \PGCgammaN{} sessions, the channel selected by $\gamma_{\mathrm{shadow}}$ is never farther in $\ell_2$ from $\alpha_X(\Lambda)$ than the measured fingerprint itself; the maximum ratio is \PGCgammaRatio{}~$\leq1$, as \cref{cor:gamma-shadow} predicts. The reconstruction and its guarantee are given in \cref{sec:concrete:gamma} of the appendix.

\subsubsection{RQ2: Conciseness and Information Retention}
\label{sec:res:rq2}


\autoref{fig:rq2} compares the size of the fingerprint interface with the full channel representation as the number of qubits increases. The compression ratio is annotated at each qubit count.

\autoref{tab:conciseness} reports fingerprint size relative to the $16^n$-dimensional channel representation. The compression ratio follows directly from the representation sizes: at $n=6$, the fingerprint is $81{,}049\times$ smaller than the full channel representation. 

The more substantive question is how much information about the tested noise parameter survives this compression. On the synthetic single-qubit distribution of \cref{sec:method:design}, the fingerprint retains nearly all information about noise strength $p$ than the full Choi matrix carries: the bootstrap mutual-information ratio is $R_{\mathrm{info}}=0.99\pm0.01$, and within-type parameter regression achieves $R^2=0.96$. These values concern recovery of $p$ on this finite distribution; they do not establish distribution-free information preservation for arbitrary channels or program outputs.

As a negative control, equal-dimension PCA embeddings and Gaussian random projections achieve the same retention ($R^2=0.96$, invariant across projection seeds). The fingerprint's advantage is therefore not generic optimality but that it preserves the exact, physically interpretable slice selected by $X$.

\subsubsection{RQ3: Measurement Efficiency}
\label{sec:res:rq3}

\begin{figure}[t]
\centering 
\begin{minipage}[c]{0.32\linewidth}
  \centering
  \includegraphics[width=1.01\linewidth,trim={0.5cm 0.5cm 0 0}, clip]{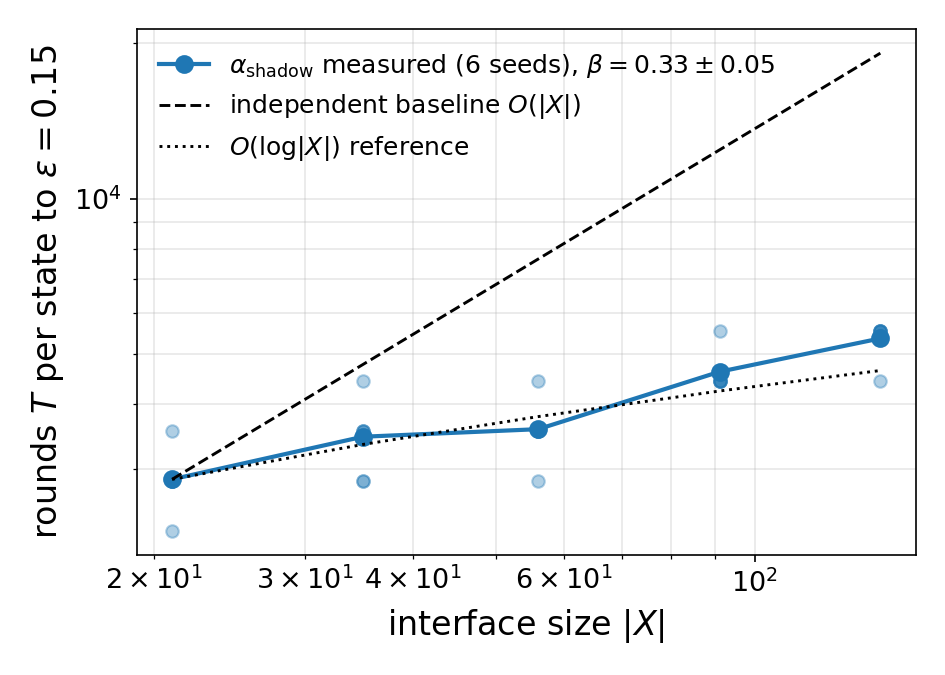}
\end{minipage}%
\hspace{0.1cm}
\begin{minipage}[c]{0.63\linewidth}
  \centering
  \includegraphics[width=1.01\linewidth]{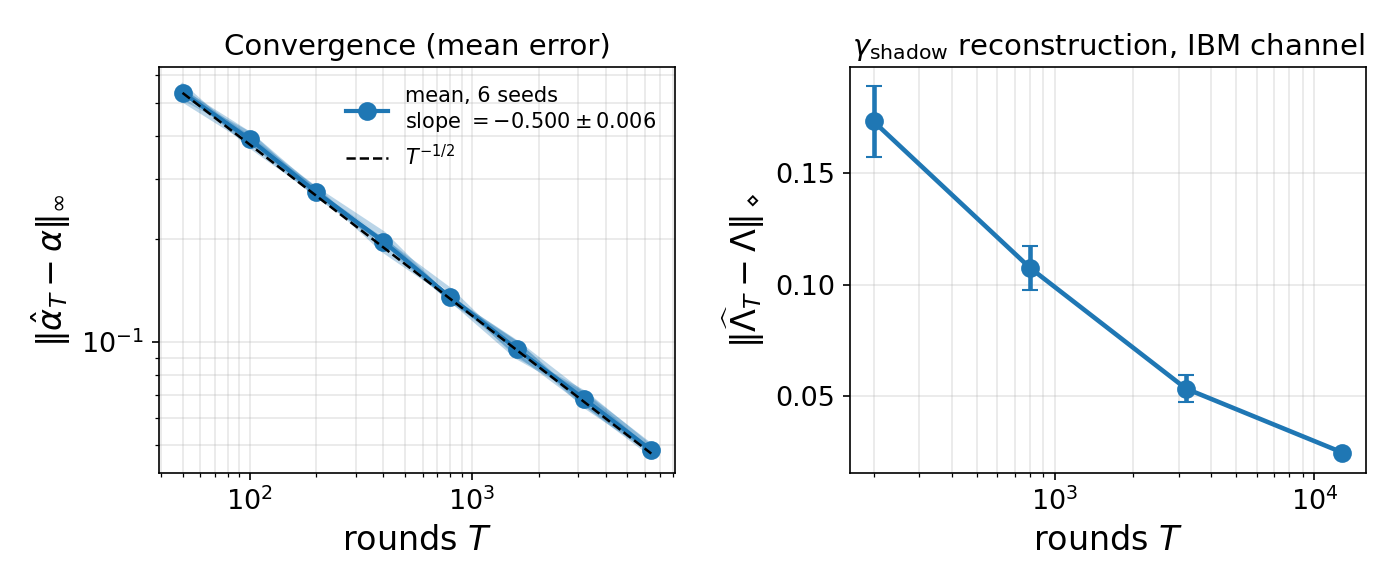}
\end{minipage}
\caption{\textbf{\textit{RQ3-RQ4's Results: }}%
\textbf{RQ3 (leftmost).} Literal-shadow rounds per state for error at most
$\varepsilon=0.15$ with confidence \ReachConf{}, vs interface size
$\lvert X\rvert$, six seeds (faint points) and their mean. The measured
growth ($\beta=\RQthreeBeta\pm\RQthreeBetaSd$) sits far below the
independent baseline ($O(\lvert X\rvert)$, dashed) and close to the
logarithmic reference (dotted).
\textbf{RQ4 (right two).} Estimator convergence at the $T^{-1/2}$ rate
(mean error, six seeds) in the mid-plot, and diamond error of the
$\gamma_{\mathrm{shadow}}$ reconstruction of a hardware-informed
IBM-Boston channel vs budget in the right plot.}
\label{fig:rq3}
\label{fig:rq4}
\Description{From left to right: a log-scale comparison of full channel
and fingerprint sizes with compression ratios annotated; a log-log plot
of measured shadow rounds vs interface size with linear and logarithmic
references; and two plots showing estimator convergence and decreasing
diamond reconstruction error as the budget increases.}
\end{figure}

\autoref{fig:rq3} (left) reports the number of rounds per reference state required to achieve uniform error below $\varepsilon=0.15$ with confidence \ReachConf{} as a function of $|X|$. Using the literal estimator of \cref{alg:alpha-shadow}, the required rounds grow substantially more slowly than the interface: as $|X|$ grows by $\RQthreeXgrowth\times$, required rounds increase by only $\RQthreeTgrowthMin{}$--$\RQthreeTgrowthMax\times$, giving fitted exponent $\beta=\RQthreeBeta\pm\RQthreeBetaSd$ (range $\RQthreeBetaMin$--$\RQthreeBetaMax$) — well below the linear baseline ($\beta=1$) and consistent with $O(\log|X|)$ theoretical rate over the tested range.

The confidence requirement increases the budget but does not steepen its growth. Under the typical-case (median) criterion, the largest interface requires $\RQthreeTmaxMed{}$ rounds rather than $\RQthreeTmax{}$, with $\beta=\RQthreeBetaMed\pm\RQthreeBetaMedSd$. Full process tomography, which scales as $16^n$, overtakes the other strategies near $n\approx3$--$4$; beyond $n=5$, tomography is intractable and the fingerprint is the only feasible option among the compared approaches. The exponent is fitted over a $\RQthreeXgrowth\times$ range in $|X|$ and is interpreted as consistent with logarithmic scaling, not as an asymptotic measurement.

\mypara{Estimator fidelity and variance}
We verify that $\alpha_{\mathrm{shadow}}$ implements the Huang--Kueng--Preskill estimator used in the analysis. Across $162$ (state, observable, channel, $n\in\{1,2\}$) cases, the maximum single-shot bias at $T=2\times10^4$ is $0.039$, consistent with zero up to Monte-Carlo error. The measured single-round variance matches the shadow-norm expression $3^{w(O)}-\Tr(O\Lambda(\rho))^2$ of eq. \ref{eq:shadow-norm} to two significant figures; representative values are reported in \autoref{tab:variance}. With snapshots shared across observables, the rounds needed for a fixed uniform error grow as $O(\log m)$ in the number of observables $m$: for weight-$1$ observables at $n=4$ and error at most $\ShotsEps$ with confidence \ReachConf{}, $m\in\{\ShotsM\}$ requires $T\in\{\ShotsT\}$ rounds (typical case: $\{\ShotsTmed\}$).

\mypara{Extended scaling to eight qubits}
The main efficiency experiment varies the number of reference states while retaining the three single-qubit Pauli observables. We additionally evaluate the literal product-factorised sampler at larger system sizes, using product reference states and weight-$\leq2$ observables (\cref{sec:method:scale} details the construction and exact ground truths). In this experiment, rounds to reach uniform error $\varepsilon=0.15$ at confidence \ReachConf{} increase from \ScaleTfirst{} to \ScaleTlast{} while the interface grows by $\ScaleXrange\times$, from $|X|=\ScaleXmin$ to $|X|=\ScaleXmax$. 
The pooled exponent is $\beta=\ScaleBeta$ (within-size exponents range $\ScaleBetaWithinMin$--$\ScaleBetaWithinMax$), providing additional evidence over a range $5.7\times$ wider than the $n\leq2$ measurement. The experiment is restricted to product instances; entangled probes are retained in the $n\leq2$ suites.

\subsubsection{RQ4: Adaptation Under Noise Drift}
\label{sec:res:rq4}

We report estimator convergence, the behaviour of optional channel reconstruction, and the effect of reusing measurements collected before a channel change (i.e., the direct test of drift-aware updating).

\mypara{Estimator convergence}
\autoref{fig:rq4} (middle) shows estimator error as a function of the number of rounds $T$ over six independent runs. The abstraction is continuous, with $\|\alpha_{t+1}-\alpha_t\|_\infty\leq2.0\,\delta_t$ in the physical drift $\delta_t$ (\cref{def:continuity}). The empirical estimator converges with log--log slope $-0.500\pm0.006$ over six seeds, matching the $O(T^{-1/2})$ rate of \cref{def:incremental}.

\mypara{Optional channel reconstruction}
\autoref{fig:rq4} (right) reports the diamond-norm reconstruction error for the general $\gamma_{\mathrm{shadow}}$ reconstruction. Applying the non-parametric concretisation of eq. \ref{eq:gamma-sdp} to a hardware-calibrated IBM channel reduces the diamond reconstruction error from $0.173\pm0.016$ at $T=200$ to $0.025\pm0.001$ at $T=12{,}800$ rounds (mean $\pm$ std, over three seeds of six repetitions each). This reconstruction is optional and not required for fingerprint comparison.

\mypara{Reusing snapshots after drift}
After a drift step with fingerprint displacement $\RQfourDrift$, pooling $1{,}600$ stale snapshots with fresh observations restores accuracy $\varepsilon=0.10$ at confidence \ReachConf{}, using $\RQfourWarm\pm\RQfourWarmSd$ fresh rounds per state versus $\RQfourCold$ for a cold start ($\approx\RQfourRatio\times$ saving; median criterion: $\RQfourWarmMed$ vs.\ $\RQfourColdMed$, $\approx\RQfourRatioMed\times$ saving).
The stale snapshots alone bring the median error close to $\varepsilon$, but their own sampling noise means that reaching $\varepsilon$ in $95\%$ of sessions still requires many fresh rounds. Stale measurements are not discarded after drift: their bias is bounded by the fingerprint drift and is diluted by pooling. The appendix further characterises the drift range over which this trade-off is beneficial.

\section{Discussion}
\label{sec:res:summary}
\label{sec:appNdiscuss}

\autoref{tab:compliance} summarises the theoretical bounds together with the corresponding empirical measurements. Every measured quantity satisfies its predicted bound, and the soundness and Lipschitz bounds are attained where the theory predicts tightness.
Although generic equal-dimension sketches retain comparable information, the fingerprint remains interpretable in terms of physically meaningful probe outcomes rather than latent coordinates.
Note that several reported quantities are mathematical consequences of the construction rather than empirical findings. We have flagged them as such: the compression ratio is a counting fact (the empirical content is retention), and the exact-soundness gap is an algebraic identity confirmed to machine precision. 
\begin{table}[t!]
  \centering
  \caption{Compliance of $\alpha_{\mathrm{shadow}}$ with the \cref{sec:properties} bounds.}
  \label{tab:compliance}
  \footnotesize
  \begin{tabular}{llll}
    \toprule
    Property (Def.)                 & Theoretical bound                      & Measured                         & Verdict \\
    \midrule
    Galois law (\cref{def:galois}) & law, Pauli channels & \GCpauliViolations{} violations in \GCpauliPairs{} pairs & holds \\
    Obs.\ soundness (\cref{prop:exact-sound}) & $D_{\mathrm{abs}}=\Delta_{\mathrm{obs}}$ & gap $1.1\!\times\!10^{-16}$      & holds (exact) \\
    Lipschitz (\cref{def:lipschitz})    & $L\le 1$                            & $L=1.00$ (AD/PD), $2/3$ (dep)    & holds, tight on AD/PD \\
    Conciseness (\cref{def:conciseness})& $\lvert X\rvert \ll 16^n$           & $81{,}049\times$, $R_{\mathrm{info}}\!\approx\!1$ & holds \\
    Efficiency (\cref{def:est-eff})     & $T=O(\varepsilon^{-2}\log\lvert X\rvert)$ & $\beta=\RQthreeBeta\pm\RQthreeBetaSd$ & holds \\
    Adaptiveness (\cref{def:incremental})& $O(T^{-1/2})$                      & slope $-0.500\pm0.006$                    & holds \\
    Robustness (\cref{def:robustness})  & $\lVert\widehat\Lambda_T-\Lambda_T\rVert_\diamond\!\to\!0$ & $0.173\!\to\!0.025$ & holds \\
    \bottomrule
  \end{tabular}
\end{table}

\subsection{Threats to validity}
Next, we discuss the main threats to validity. The appendix (\cref{sec:appendix}) reports additional experiments extending the evaluation to larger systems and hardware-inspired settings.

\mypara{\textbf{Internal validity}}
The reference implementation was validated against the formal development by encoding and testing every definition and proposition, confirming that stated (in)equalities hold during execution and reducing the risk that findings are implementation artefacts (see also Appendix \cref{sec:theory2Emp}). All stochastic quantities were replicated over five independent seeds (reported mean~$\pm$~std); the fine-grid protocol of \cref{sec:method:design} supersedes an earlier coarser budget quantisation.
\mypara{\textbf{Construct validity}}
The concrete-domain limitations are measured ~\cref{sec:res:limits}: the detectability and readout floors, the first-order blindness to coherent $ZZ$ crosstalk, the blind subspace induced by finite experiment interfaces, and the drift region in which warm-starting remains beneficial. Expert signatures are currently hand-specified for a two-qubit suite; scaling their verification to larger $n$ requires generating signatures from the channel's Pauli-transfer-matrix action, which we leave to future work.

\mypara{\textbf{External validity}} 
The efficiency exponent is fit over a $6.7\times$ range in $\lvert X\rvert$ and is consistent with, not a proof of, logarithmic scaling. All channels are synthetic or IBM-calibrated rather than live device data; the \texttt{ibm\_boston} profile uses the platform's published medians for two-qubit and readout error, with Heron-class constants for $T_1$, $T_2$, and single-qubit error. An on-device comparison is left to future work.
Robustness under realistic conditions was assessed via a configurable noise framework and stress protocols covering detectability, non-Pauli errors, blindness, and drift validity; results are reported in Appendix \cref{sec:method:limits}.
The limitations discussed below (\cref{sec:framework:limitations}) concern the framework itself rather than this particular instantiation.

\begin{wrapfigure}{r}{0.37\textwidth} 
\centering
\includegraphics[width=0.93\linewidth]{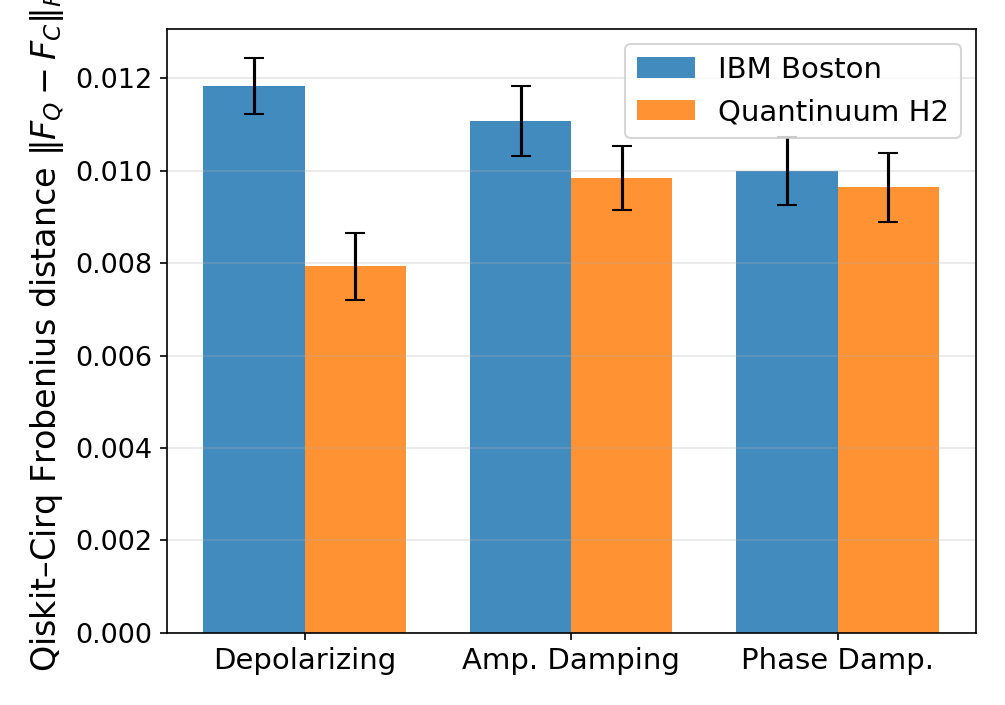}
\caption{Cross-platform fingerprint distances between two simulator backends per noise family, with bootstrap standard errors.}
\label{fig:crossplatform}
\Description{Bar chart of cross-platform fingerprint distances for three noise families under two hardware profiles, with error bars.}
\end{wrapfigure}
\mypara{\textbf{Generalisability}}%
We assessed scalability (w.r.t. the number of qubits) by introducing scalable ground truths and a product-factorised literal sampler; we report constructions and results in Appendix \cref{sec:scale}.
We evaluated discriminability beyond the channels used in the main evaluation (RQ1-RQ4) via Frobenius distances within and between noise families and across platforms. The mean within-family fingerprint distance is $0.0009$ versus $0.7806$ between families ($876\times$ separation), confirming that distinct noise mechanisms occupy well-separated regions of $\mathcal{F}$. 

Cross-platform distances between two independent simulator backends (Qiskit and Cirq), on IBM Boston and Quantinuum H2 profiles (\autoref{fig:crossplatform}), range from approximately $0.008$ to $0.012$ across the three noise families. The bootstrap standard errors are at most $0.0008$ (minimum distance-to-uncertainty ratio $11\times$), confirming that the residual platform discrepancy is measurable rather than estimator noise. Fingerprints thus support both noise-family attribution and platform discrepancy analysis.



\subsection{Limitations}
\label{sec:framework:limitations}

The abstraction's discriminative power depends on the chosen experimental interface. While the framework does not mandate a single canonical interface, it specifies the properties any admissible one must satisfy (soundness, conciseness, efficiency, adaptiveness), and the interface should be selected to match the goals of the intended analysis.

Since the abstraction is inferred from measurement outcomes rather than derived from a channel's mathematical description, it is applicable to real hardware where the exact model is unknown. However, the resulting fingerprint retains dependence on the chosen interface and the finite measurement budget, so repeated evaluations of the same channel may yield slightly different fingerprints; the soundness definitions of \cref{subsec:soundness} explicitly account for this variability.


The $(\varepsilon,\delta)$ guarantee of \cref{def:est-eff} bounds the probability that a measured fingerprint deviates from its exact interface profile by more than $\varepsilon$. For the fixed four-prefix illustration, a union bound converts profile-level coverage into simultaneous distance intervals (Appendix, \cref{app:diff-debug}). This does not extend to comparing full measurement distributions across environments or to end-to-end debugging workflows with data-dependent checkpoint selection, compiler alignment, and total acquisition cost; formalising such an extended contract remains future work.


Finally, the expert validation uses observables containing identity gates (\eg{} $ZI$, $IX$), whose meaning depends on how idle qubits are realised by the execution platform. In our evaluation, channels are represented via Kraus operators and such observables are interpreted algebraically as tensor-product operators, so the reported results are valid under that representation. In realistic execution settings, such as those in Appendix \cref{sec:method:limits}, this distinction becomes relevant, and extending the evaluation methodology accordingly is left to future work.
\section{Related Work}
\label{sec:related-work}

\mypara{Characterising a device versus answering an analysis query}
Quantum process tomography, gate-set tomography, and randomised benchmarking characterise channels or summarise error rates~\cite{chuang1997prescription,Blume-Kohout2017,magesan2011scalable}, but answer a different question from ours: what can a particular analysis observe about two execution environments? A complete channel description answers arbitrary such queries at high cost; a benchmark statistic is cheaper but need not distinguish the behaviours selected by an analysis. Our fingerprint records the values of a declared, finite experiment interface, and its guarantee is \emph{relative to that interface}.
%
Classical shadows reduce the measurements needed to estimate many observable properties~\cite{10.1145/3188745.3188802,10.1145/3313276.3316378,huang2020predicting}, and have been applied to process tomography via Choi-representation estimation~\cite{levy2024shadowqpt}. We use the classical-shadow estimator for a narrower, analysis-directed purpose: the interface is fixed before acquisition, and one shared measurement record estimates all coordinates with an explicit finite-shot confidence bound.


\mypara{Noise-aware testing and empirical fingerprints}
Device ``noise fingerprints'' have been used for identification and drift tracking~\cite{martina2022learning}. The closest precursor introduced live fingerprints to compare noisy simulators~\cite{simshadow:ICST:2026}. We build on this by making the abstraction boundary explicit: equality and distance mean agreement and discrepancy on the selected observations, interface extension refines what can be distinguished, and finite-shot estimation provides probabilistic coverage guarantees. Our implementation also reuses a shared randomized measurement record across observables instead of collecting separate batches per coordinate.
Related work targets complementary aims: noise-aware property-based testing reduces spurious assertion failures~\cite{pontolillo2025,10682972}, simulator fuzzing searches for behavioural inconsistencies~\cite{fuzzq}, and delta debugging identifies sources of quantum regressions~\cite{pontolillo2024delta}. Our prefix example highlights the first \emph{observed} difference between configurations but is not a general fault-localization method.

\mypara{Abstract interpretation and uncertainty}
Classical and quantum abstract interpretations provide a language for relating tractable representations to concrete behaviour~\cite{Cousot77,10.1007/978-3-540-69166-2_18,10.1145/3453483.3454061,FENG2023105077}, supporting program-state analysis and correctness.
Here, the concrete object is the noisy channel induced by the execution environment, and the abstractions are observations tailored to a software-analysis task. Probabilistic abstract interpretation already addresses reasoning over quantitative and uncertain systems~\cite{goos_backwards_2001,hutchison_probabilistic_2012}.
Our probabilistic Galois connection follows the order-theoretic framework of probabilistic abstract interpretation: concretisation yields a distribution over channels, compared by expected value (\cref{def:galois}), using orders from Coecke and Martin~\cite{Coecke2011}.
Finally, we distinguish two limits: the $(\varepsilon,\delta)$ bound of \cref{def:est-eff} controls measurement error for a fixed channel, whereas information loss outside the interface is an intentional design choice, not a finite-sampling artefact.

\section{Conclusion and Future Research Avenues}
\label{sec:conclusion}

We presented noise fingerprints as an interface-relative semantic abstraction for noisy quantum executions, separating the concrete semantics of a noisy channel from the finite observational interface through which a programmer, testing tool, or analysis is observed. A fingerprint is thus not merely an empirical summary but the abstract denotation of a noisy execution with respect to a chosen set of reference states and observables. The abstraction yields a formal reasoning framework: exact fingerprints are complete for the chosen interface, empirical fingerprints provide statistically controlled approximations, and interface refinement monotonically increases observational precision. Classical shadows enable efficient implementation, delivering compact behavioural summaries without the cost of full process tomography.

\section*{AI Use}
Claude (Anthropic) helped reorganise the existing replication package and check that the manuscript’s claims matched the reported results.
Claude and ChatGPT (OpenAI) assisted with manuscript clarity and the LaTeX formatting of formal definitions.
The authors reviewed the outputs. Neither tool implemented \tool{}, generated or modified data, ran simulations, or performed statistical analyses.

\section*{Data-Availability Statement}

An anonymised artifact accompanies this submission. It contains the
ChanShadow reference implementation, the notebooks and scripts used
to conduct the experiments, fixed dependency versions, random seeds,
a containerised execution environment, and an automated test suite.
The artifact is intended to support reproduction of the channel-level
simulations, fingerprint estimation experiments, and supplementary
evaluations reported in this paper.

The evaluation uses simulated quantum channels and hardware-inspired
noise profiles derived from published calibration information; it
does not include measurements collected directly from physical
quantum hardware. Consequently, the artifact reproduces the
simulation-based results and does not provide an on-device validation
of the proposed approach.

The anonymised artifact is available at 
\url{https://zenodo.org/records/23414157}.

The authors intend to submit the artifact for the PLDI 2027
Artifact Evaluation process.

\bibliographystyle{ACM-Reference-Format}
\bibliography{refs}

\appendix

\newpage
\section{Appendix}
\label{sec:appendix}

This appendix collects material that supports the formal development and main evaluation without interrupting their presentation. It includes the worked program-level example (\cref{app:diff-debug}), probability and quantum-computing background, the optional Choi-matrix reconstruction procedure, extended scaling experiments, and hardware-inspired stress tests. The additional experiments characterise both the regimes in which the interface is informative and the limitations that arise when relevant behaviour is not represented by the selected probes.

\subsection{Worked Examples 2-5: Noise-Aware Analysis with \tool{}}
\label{app:diff-debug}

The artifact includes a Jupyter notebook that uses noise fingerprints to check compiler optimisations, assess test portability, and locate observable differences during debugging. To try the analyses on another program, set \texttt{MY\_PROGRAM} to its source code in OpenQASM format. We briefly describe all three analyses in the \href{https://zenodo.org/records/23414157}{anonymised notebook} before presenting here the noise-aware differential debugging in detail.

Each example could form the basis of a follow-up paper building on the formalism developed here, which we leave to future work.

\paragraph{Compiler Optimisation.}
\emph{Is this pass semantics-preserving on this machine?} An optimisation that preserves ideal semantics can change observable behaviour under noise. We compare the fingerprints of a baseline execution and a candidate execution that models the effect of an optimisation. If their distance exceeds the analysis tolerance, the tool emits a warning identifying the input and observable with the largest discrepancy. 

See Section~5 of \code{worked\_example.ipynb} in the \href{https://zenodo.org/records/23414157}{anonymised artefact} for details and instructions on running this part.

\paragraph{Test Portability.}
\emph{A test suite was validated against one realisation. Can it be trusted on a new backend or after calibration drifts?} We use the suite's inputs and observables to compare executions across environments. The fingerprint distance gives the largest change in the quantities the suite observes. For the Bell program, tests of population miss the phase rotation, which measurements in the superposition bases can reveal.

See Sections~5-6 of \code{worked\_example.ipynb} in the \href{https://zenodo.org/records/23414157}{anonymised artefact} for details and instructions on running this part.

\paragraph{Debugging.}
\emph{A test fails on one realisation and passes on another. Which gate should we investigate?} We compare fingerprints at successive program prefixes and report the first prefix whose discrepancy exceeds the tolerance. In the miscalibrated-gate example, the discrepancy first appears after the first $CX$, giving the developer a concrete place to investigate.

See Section~7 of \code{worked\_example.ipynb} in the \href{https://zenodo.org/records/23414157}{anonymised artefact} for details and instructions on running this part.

\paragraph{Noise-Aware Differential Debugging.}
This section presents a supplementary program-level worked example illustrating how a finite observation interface can guide inspection of two noisy realisations of the same program. Unlike the two-qubit motivation in \Cref{sec:motivation}, this example uses a three-qubit program with four operations and examines the observed discrepancy at successive program prefixes. Its purpose is to illustrate a possible debugging workflow and, in particular, the distinction between the first observed threshold crossing and causal fault localisation.

\mypara{Program and concrete realisations}
The three-qubit source program is
\[
P=H(q_0);\;CX(q_0,q_1);\;T(q_2);\;CX(q_1,q_2).
\]
Write $P_j$ for its first $j$ gates and $\Lambda_{b,j}$ for the input-to-output channel realised by $P_j$ in configuration $b\in\{\textsf{A},\textsf{D}\}$. Thus $\Lambda_{b,j}$, not the gate name or the source text, is the concrete object compared at checkpoint $j$. The ideal prefix has unitary $U_j$, the same in both configurations. The notebook represents $\Lambda_{b,j}$ as a $4^3\times4^3$ superoperator, containing $16^3=4096$ matrix entries.

Both configurations apply one-qubit depolarising noise with rate $p_1=0.020$ after a one-qubit gate, followed by amplitude damping with rate $\gamma=0.01$ on that gate's target. After a $CX$, they apply the same sequence independently to each participating qubit, but with a two-qubit-gate depolarising rate $p_2=0.05$ in \textsf{A} and $p_2=0.09$ in \textsf{D}. These are the notebook's illustrative gate-noise parameters, not a calibration of a named physical backend. Its depolarising and damping channels are specified by
\[
\begin{aligned}
\mathcal D_p(\rho)&=\left(1-\frac{3p}{4}\right)\rho
 +\frac p4\sum_{Q\in\{X,Y,Z\}}Q\rho Q,\\
K_0&=\begin{pmatrix}1&0\\0&\sqrt{1-\gamma}\end{pmatrix},
\qquad K_1=\begin{pmatrix}0&\sqrt\gamma\\0&0\end{pmatrix},\\
\mathcal A_\gamma(\rho)&=\sum_{r=0}^1 K_r\rho K_r^\dagger.
\end{aligned}
\]
For each gate, the realised channel first applies its ideal unitary and then the applicable $\mathcal A_\gamma\circ\mathcal D_p$ on its targets. Prefixes are built by composing these operations in source order. The configurations are identical through $H(q_0)$ and $T(q_2)$ as isolated gates; the controlled change is the noise rate attached to each $CX$.

\mypara{The diagnostic interface}
Each prefix starts afresh from $\rho_0=\ket{000}\!\bra{000}$. Let $\mathcal O_{\rm diag}$ contain every non-identity three-qubit Pauli string of weight at most two, and use the same diagnostic interface $X=\{\rho_0\}\times\mathcal O_{\rm diag}$ at every prefix. There are $\binom{3}{1}3=9$ weight-one strings and $\binom{3}{2}3^2=27$ weight-two strings, hence $|X|=36$. Examples include $XII$, $ZZI$, and $IYZ$; $XXX$ is \emph{not} in this diagnostic suite because its weight is three. Keeping the suite fixed makes the four distances comparable. It is broader in number of checks than the finished program's three-check test interface, $\{ZZI,IZZ,XXX\}$, used elsewhere in the notebook, but it does not contain that interface.

For $O\in\mathcal O_{\rm diag}$, the exact fingerprint entry is the deviation from the ideal prefix,
\[
F_{b,j}(O)=\operatorname{tr}\!\left[O\Lambda_{b,j}(\rho_0)\right]
 -\operatorname{tr}\!\left[O U_j\rho_0U_j^\dagger\right].
\]
The comparison is the uniform distance
\begin{equation}\label{eq:appendix-prefix-distance}
d_j=\|F_{\textsf{A},j}-F_{\textsf{D},j}\|_\infty
 =\max_{O\in\mathcal O_{\rm diag}}\left|
 \operatorname{tr}\!\left[O\Lambda_{\textsf{A},j}(\rho_0)\right]
 -\operatorname{tr}\!\left[O\Lambda_{\textsf{D},j}(\rho_0)\right]
 \right|.
\end{equation}
The common ideal term cancels in the second expression. Thus $d_j$ is an exact maximum discrepancy for the \emph{selected} observations, not a probability that gate $j$ has failed and not a distance between full channels.

\mypara{Reading the prefix trace}
The notebook evaluates \cref{eq:appendix-prefix-distance} on all four truncated programs:
\begin{center}
\small
\begin{tabular}{@{}clc@{}}
\toprule
Prefix & Last gate & $d_j$ \\
\midrule
$P_1$ & $H(q_0)$ & $0.0000$ \\
$P_2$ & $CX(q_0,q_1)$ & $0.0729$ \\
$P_3$ & $T(q_2)$ & $0.0729$ \\
$P_4$ & $CX(q_1,q_2)$ & $0.1007$ \\
\bottomrule
\end{tabular}
\end{center}
Here $\tau$ is a developer-specified acceptable change in an expectation value. It is \emph{not} the estimator error $\varepsilon$ of \cref{def:est-eff}. The first crossing is $P_2$: the two configurations agree on the diagnostic suite after $H$, and differ beyond $\tau$ after the first $CX$. The controlled model explains this particular result: \textsf{D} increases only the error rate used after two-qubit gates. The unchanged rounded maximum at $P_3$ does not establish that $T$ has no effect on every coordinate. Likewise, the rise at $P_4$ does not by itself attribute a separate fault to the second $CX$.

The notebook's final-output test instead reads $ZZI$, $IZZ$, and $XXX$, and reports distance $0.1231$ on $P_4$. This need not equal the diagnostic value $0.1007$: $XXX$ is visible to the test but absent from the weight-at-most-two diagnostic suite. The difference illustrates both the usefulness and the limits of an interface. Neither number is a distribution-free measure of overall program equivalence.

\mypara{What ``first'' means operationally}
The notebook truncates the parsed gate sequence and composes the corresponding ideal and noisy matrices for each $P_j$. It does not interrupt one execution at successive gates, and it does not exercise an optimising compiler. On a device, each prefix would need to be prepared and run as a separate circuit; an optimising compiler might then fuse or reorder gates differently from the full program. A deployed analysis would need to specify how source checkpoints align with the realised operations before treating $P_2$ as a location in compiled code. In any setting, an earlier difference invisible to $X_1$ could first become visible in $X_2$. ``First divergence'' therefore means the first \emph{observed threshold crossing}, not a general causal fault-localisation result. A linear scan finds that crossing in this example; bisection would require a monotonicity property that these distances do not generally have.

The immediate developer action is to inspect the first $CX$ implementation and the state it receives, then evaluate a candidate change. For example, $CX$ is ideally equivalent to $H$ on the target, followed by $CZ$, followed by another $H$ on the target. Whether that replacement improves the \emph{noisy} realisation must be measured; the example does not establish that it does.

\mypara{If the profiles are measured instead}
The trace above is analytic. For finite-shot fingerprints $\widehat F_{b,j}$, let $\varepsilon_s$ bound the uniform error of \emph{each} profile at failure probability at most $\delta/8$. There are two configurations at four checkpoints, so a union bound gives, simultaneously for every $j$, with probability at least $1-\delta$,
\[
\bigl|\widehat d_j-d_j\bigr|\le 2\varepsilon_s,
\qquad
\widehat d_j=\|\widehat F_{\textsf{A},j}-\widehat F_{\textsf{D},j}\|_\infty.
\]
A confidence-aware claim that $P_2$ is the first crossing of $\tau$ would require $\widehat d_1+2\varepsilon_s\le\tau$ and $\widehat d_2-2\varepsilon_s>\tau$; otherwise the result is undecided at that confidence. The notebook supplies no such finite-shot experiment, so its four values should not be read as certified detections. With $T$ measurement rounds for each profile, a direct four-prefix comparison would require eight profile acquisitions (and thus $8T$ circuit executions under that allocation); caching the baseline would leave four new acquisitions for \textsf{D}. Compilation, preparation, readout, and any calibration overhead are additional costs. The channel-level evaluation in the paper does not measure this complete debugging workflow.

See \code{worked\_example\_Part\_4\_noise\_aware\_diff\_testing.ipynb} in the \href{https://zenodo.org/records/23414157}{anonymised artefact} for details and instructions on running this part. This is the only example with a fixed program.

\subsection{Quantum Computing Background}
\label{subsec:background}
For a detailed background, see \cite{nielsen2010quantum}. 
A (pure) quantum state is represented by a ket $\vert \psi \rangle$ in a complex Hilbert space. 
For a single \textbf{\textit{qubit}}, the fundamental unit of quantum information, in the computational basis $\{\vert 0 \rangle, \vert 1 \rangle\}$, the general state is a \textit{superposition} of the basis states:
{
    \setlength{\abovedisplayskip}{1pt}
    \setlength{\belowdisplayskip}{1pt}
    \setlength{\abovedisplayshortskip}{1pt}
    \setlength{\belowdisplayshortskip}{1pt}
    \small
    \begin{equation} \label{eqPsi}
    \vert \psi \rangle = \alpha \vert 0 \rangle + \beta \vert 1 \rangle,
    \end{equation}
    \normalsize
}%
where \( \alpha \) and \( \beta \) are complex numbers called \textit{amplitude coefficients}, encoding both the likelihood of observing each state and the \textit{phase} of the quantum system (\eg{} 
$\ket{+}=(\ket{0}+\ket{1})/\sqrt{2}$ and $\ket{+i}=(\ket{0}+i\ket{1})/\sqrt{2}$). 
The product state of an $n$ qubit system can be described by 
$\vert \psi \rangle = \vert \psi_1 \rangle \otimes \vert \psi_2 \rangle \otimes \cdots \otimes \vert \psi_n \rangle$, where $\otimes$ denotes the tensor product and each $\psi_i$ ($1 \le i \le n$) is a single qubit state (\cref{eqPsi}). The associated Hilbert space is $ 2^n$-dimensional, which grows exponentially with $n$, and contains the quantum states and quantum-channel representations. 

\textbf{\textit{Entanglement}} occurs when two or more qubits are correlated thus, the state of one qubit cannot be described independently of the others.
An entangled system is described by a single joint quantum state that is non-separable. It can neither be factored into a simple tensor product (with~$\otimes$) of individual qubit states, nor as a probabilistic mixture of such product states. This is essential for capturing the non-classical correlations between the qubits. E.g. the \textit{\textbf{Bell states}} \cite{nielsen2010quantum}, a set of four maximally entangled two-qubit states,
{
\setlength{\abovedisplayskip}{1pt}
\setlength{\belowdisplayskip}{1pt}
\setlength{\abovedisplayshortskip}{1pt}
\setlength{\belowdisplayshortskip}{1pt}
\small
\begin{equation}
    |\Phi^{\pm}\rangle = \frac{1}{\sqrt{2}} (|00\rangle \pm |11\rangle) \text{,        } \qquad 
    |\Psi^{\pm}\rangle = \frac{1}{\sqrt{2}} (|01\rangle \pm |10\rangle),
\end{equation}
\normalsize}%
have been used as part of the reference set in constructing our instantiation of a fingerprint abstraction (\cref{sec:concrete}).
The most general quantum state of an \(n\)-qubit system is described by a density matrix $\rho \in \mathbb{C}^{2^n \times 2^n},$ representing either a \textit{pure state} with $\rho = |\psi\rangle \langle \psi|$, 
or a \textit{mixed state} with $\rho = \sum_i p_i |\psi_i\rangle \langle \psi_i|$, scaling exponentially with \(n\).

\subsubsection{Quantum operators}
\label{sec:apx:qops}
\textit{Quantum operators}, or \textit{gates}, \eg{} NOT or rotation gates, are linear transformations acting on a qubit or several qubits, and are represented by \textit{unitary matrices}.
By definition, they are \textit{reversible} operations such that each matrix has an inverse that undoes its action.
A matrix \(U\) is unitary if:
$UU^\dagger = U^\dagger U = I$, 
where \(U^\dagger\) is the \textit{Hermitian conjugate} (or adjoint) of \(U\), and \(I\) is the identity matrix.
The \textit{Hermitian conjugate} of a matrix, \(A = \{a_{ij}\}\), is obtained by taking the transpose and complex conjugate of each entry: $A^\dagger = \overline{\{a_{ji}\}}$. A matrix \textit{A}
is \textit{Hermitian} if it satisfies: $A = A^\dagger$, meaning it is equal to the transpose of its own complex conjugate and hence is symmetric.

A measurable outcome, or an \textbf{\emph{observable}} $O$, is represented by a Hermitian matrix whose eigenvalues are the possible outcomes of the measurement. The expected result of measuring $O$ on a system~$\vert \psi \rangle$ is $\langle \psi \vert O \vert \psi \rangle$.
Our instantiation of a fingerprint abstraction (\cref{sec:concrete}) uses two-qubit Pauli matrices\footnote{ 
We give here the definition of Pauli matrices, but their exact definition is not required for understanding this work; 
{
    \setlength{\abovedisplayskip}{1pt}
    \setlength{\belowdisplayskip}{1pt}
    \setlength{\abovedisplayshortskip}{1pt}
    \setlength{\belowdisplayshortskip}{1pt}
    \small
    \begin{equation*}
    \text{\Rxgate{}}\qquad\text{\Rygate{}}\qquad\text{\Rgate{}.}
    \end{equation*}
    \normalsize
}%
}
and other quantum operators as part of the reference set.

\subsubsection{Execution of Quantum Programs in Practice}
\label{sec:apx:shots}
\textbf{\textit{"Shots" in Quantum Experiments.}} %
In quantum computing, a single circuit execution yields only one measurement outcome. Estimating a probability distribution, therefore, requires many repeated circuit executions, or \emph{shots}, with the standard error scaling as $1/\sqrt{\text{N}}$ for N independent shots.
Further, an entire experiment of many shots may itself be repeated multiple times (\emph{repeats}) to increase total sample size, reduce statistical error and enable checking for variability across runs.
We utilise both concepts in our work. 
\emph{Shots} are directly associated with our instantiation of a fingerprint abstraction and methodology, while \emph{repeats} are part of the evaluation methodology.

We model noise using \textbf{\emph{quantum channels}}, which are \textit{Completely Positive, Trace-Preserving} (CPTP) \textit{maps} acting on density matrices. A quantum channel can be viewed as the mathematical description of the physical process through which a quantum system is transferred from a sender to a receiver. Interactions between the quantum system and its surrounding environment that alter the transmitted state are modelled in a noisy channel. The channel is considered \textit{noiseless} when the quantum state remains undisturbed during transmission. A channel $\Lambda$ acting on $k$ qubits can be expressed in the following form, known as the \textbf{\textit{Kraus form}}:
{
    \setlength{\abovedisplayskip}{1pt}
    \setlength{\belowdisplayskip}{1pt}
    \setlength{\abovedisplayshortskip}{1pt}
    \setlength{\belowdisplayshortskip}{1pt}
    \small
\begin{equation}
\label{eq:kraus}
\Lambda(\rho) = \sum_\ell K_\ell \rho K_\ell^\dagger \texttt{ with } \sum_\ell K_\ell^\dagger K_\ell = I.     
\end{equation}
\normalsize
}%
This representation guarantees that the evolution of density matrices remains physically valid, even when extended to larger composite systems. We use the Kraus form in \cref{sec:methodology} and \cref{sec:results}.

Among the most commonly used noise channels are the \emph{depolarising}, \emph{amplitude damping}, and \emph{phase damping} channels. \textbf{\textit{Depolarising noise}} models random errors that can drive the quantum state towards the maximally mixed state by contracting the Bloch vector. \textbf{\textit{Amplitude damping}} captures energy dissipation phenomena, such as relaxation from the excited state $\lvert 1 \rangle$ to the ground state $\lvert 0 \rangle$, while \textbf{\textit{phase damping}} models the loss of quantum coherence without altering measurement probabilities in the computational basis. These channels provide mathematical abstractions of physical noise processes and are essential for developing, testing, and evaluating quantum programs under realistic noisy conditions. We utilise these channels in \cref{sec:methodology}, \cref{sec:results} and \cref{sec:appNdiscuss}.

\subsubsection{Classical Shadows in a Nutshell}
\label{subsec:classical-shadows}
Quantum computations are affected by noise from imperfect control, decoherence, and measurement errors \cite{nielsen2010quantum}. Noise is studied on \textit{quantum hardware}, where device-specific effects can be time-dependent, correlated, and non-Markovian, or in \textit{simulation}, where errors are modelled via \textit{abstract quantum channels} (\eg{} depolarising or amplitude-damping \cite{cirq_representing_noise_2025}) but may fail to reproduce accurately the statistical behaviour observed on devices.  \textit{Error correction} is a set of methods used to address the loss of quantum information due to noise and decoherence, enabling errors to be detected and corrected such that no information about the encoded logical state is revealed.

\textit{Error mitigation} reduces noise impact without the full overhead of fault-tolerant quantum \textit{error-correction}, typically via classical post-processing or circuit-level transformations \cite{temme2017error,endo2018practical}. \emph{Classical shadows}~\cite{huang2020predicting} are a common complementary randomised-measurement technique, within the broader family of shadow tomography~\cite{10.1145/3188745.3188802,10.1145/3313276.3316378} that estimate many observable expectation values without reconstructing the full quantum state. 

Classical shadows are a specific efficient instantiation generated by applying randomly chosen unitaries, $U_t$, to an unknown quantum state, $\rho$, measured in the computational basis, and then post-processing the outcomes into snapshot operators, $\hat{\rho}_t$, such that $\mathbb{E}[\hat{\rho}_t] = \rho$.
The resulting multiset, ${\hat{\rho}_t}$, allows unbiased simultaneous estimation of many expectation values, say,  $\Tr(O\rho)$ for each observable $O$ in state $\rho$  for observables $O$ as $\widehat{\langle O \rangle} = \frac{1}{T}\sum_{t=1}^T \Tr(O \hat{\rho}_t).$

Randomising measurement bases ensures that each measurement outcome contains information about multiple observables. Importantly, if observables have a bounded shadow norm, the number of measurements scales logarithmically rather than linearly with the number of observables. This approach has been utilised in physical applications, \eg{} characterising laboratory-prepared states.

The construction of our instantiation of fingerprint abstraction, \tool, as detailed in \cref{sec:concrete}, is inspired by the classical-shadow framework of Huang et al.~\cite{huang2020predicting}. However, we do not materialise full snapshots. 
\tool{}, instead, maps measurement outcomes directly to deviation signatures tailored to QSE.
Specifically, we restrict the measurement ensemble to a fixed, finite set of observables relevant for quantum-software workloads: per reference state $\rho_i$ and observable $O_j$, we collect an empirical estimator $\widehat{\langle O_j \rangle}_i$ and standard error $\sigma_{ij}$.

\subsection{Probability Preliminaries}
\label{sec:app:prob}
This section recalls the notions used in \cref{def:galois,def:est-eff,prop:pgc-tolerance,cor:pgc-shadow} and in the evaluation of RQ1.

\mypara{Measurable sets and distributions}
A $\sigma$-algebra on a set $Y$ is a family of subsets of $Y$ that contains $Y$ and is closed under complement and countable union; its members are the events. On a subset of a finite-dimensional normed vector space we use the Borel $\sigma$-algebra, generated by the open sets. A probability distribution on $Y$ is a function $P$ from events to $[0,1]$ with $P(Y)=1$ that is additive over countable unions of disjoint events. It need not have finite support: a distribution over channels may spread over a continuum. The point mass $[y]$ assigns $1$ to every event that contains $y$ and $0$ to the others.

\mypara{Expectation}
If $Y$ is a subset of a finite-dimensional real vector space and $P$ has a finite first moment, its expectation is $\mathbb{E}[P]=\int_Y y\,dP(y)$, computed coordinate by coordinate; for a point mass, $\mathbb{E}[[y]]=y$, and for $P=\sum_iw_i[y_i]$, $\mathbb{E}[P]=\sum_iw_iy_i$. If $Y$ is closed and convex, $\mathbb{E}[P]$ lies in $Y$, because it lies in the closed convex hull of the support of $P$. The channels on $n$ qubits form a compact convex set, so every distribution over channels has an expectation, and that expectation is again a channel; this is the expected concretisation $\bar\gamma_P$ of \cref{def:galois}.

\mypara{Random fingerprints, laws and quantiles}
A measured fingerprint is a random variable $\widehat F$ with values in $\mathbb{R}^X$; its law is the distribution $P(B)=\Pr[\widehat F\in B]$, which is $\alpha_T(\Lambda)$ in \cref{def:est-eff}. For a real random variable $D$, such as the error $D_\Lambda=\|\alpha(\Lambda)-\widehat F\|_\infty$, the function $t\mapsto\Pr[D\le t]$ is non-decreasing and right-continuous, and the $(1-\delta)$-quantile of $D$ is $q_{1-\delta}=\inf\{t:\Pr[D\le t]\ge1-\delta\}$. By right-continuity, $\Pr[D\le t]\ge1-\delta$ holds exactly when $t\ge q_{1-\delta}$, which is \cref{prop:pgc-tolerance}(2).

\mypara{Repeated sessions and empirical coverage}
Independent repetitions of a measurement session give independent random fingerprints with a common law. Over $S$ sessions, the number of sessions with $D\le\varepsilon$ is binomially distributed with parameter $c=\Pr[D\le\varepsilon]$; its relative frequency estimates the coverage $c$, and the Clopper--Pearson bound is an exact one-sided lower confidence bound for $c$. The empirical $(1-\delta)$-quantile of $S$ sessions is the $\lceil(1-\delta)S\rceil$-th smallest observed error.

\mypara{Concentration}
The union bound, $\Pr[\bigcup_iA_i]\le\sum_i\Pr[A_i]$, turns guarantees for single entries into a guarantee for the whole fingerprint. Hoeffding's inequality states that the mean $\bar Y$ of $T$ independent random variables with values in $[a,b]$ satisfies $\Pr[|\bar Y-\mathbb{E}\bar Y|\ge\varepsilon]\le2\exp(-2T\varepsilon^2/(b-a)^2)$. Chebyshev's inequality states that $\Pr[|Y-\mathbb{E}Y|\ge\varepsilon]\le\mathrm{Var}(Y)/\varepsilon^2$. The median-of-means estimator combines the two: each batch mean is $\varepsilon$-accurate with probability at least $33/34$ once a batch has $34\,\mathrm{Var}(Y)/\varepsilon^2$ rounds, and a Chernoff bound on the number of inaccurate batches shows that $K=\lceil2\ln(2M/\delta)\rceil$ batches make all $M$ estimates $\varepsilon$-accurate with probability at least $1-\delta$~\cite{huang2020predicting}.

\subsection{Optional Channel Reconstruction}
\label{sec:concrete:gamma}

A fingerprint constrains the underlying channel only through the $|X|$ linear functionals
\[
  \Tr\!\left(O_j\Lambda(\rho_i)\right)
  =\Tr\!\left[(\rho_i^{\mathsf T}\otimes O_j)J_\Lambda\right],
\]
where $J_\Lambda$ is the Choi matrix in the input-first convention. When a concrete channel representative is required, we define the partial concretisation $\gamma_{\mathrm{shadow}}$ by the physically constrained least-squares problem
\begin{equation}
\gamma_{\mathrm{shadow}}(\widehat F)
=
\operatorname*{arg\,min}_{J\succeq0,\;\Tr_{\mathrm{out}}(J)=I}
\sum_{(\rho_i,O_j)\in X}
\Bigl(
\Tr\!\left[(\rho_i^{\mathsf T}\!\otimes O_j)J\right]
-
\bigl(\widehat F_{ij}+\mu^{\mathrm{ideal}}_{ij}\bigr)
\Bigr)^2,
\label{eq:gamma-sdp}
\end{equation}
which is a semidefinite program. The constraints $J\succeq0$ and $\Tr_{\mathrm{out}}(J)=I$ enforce complete positivity and trace preservation, respectively. When $X$ is informationally complete for the channel family of interest, the optimisation recovers $\Lambda$ exactly from the noiseless fingerprint. Otherwise, it returns a least-squares CPTP map consistent with the observed deviations. Unlike the parametric reconstruction used in the proof of concept, this formulation is non-parametric and applies to arbitrary CPTP channels.

\begin{corollary}[Precision of $\gamma_{\mathrm{shadow}}$]\label{cor:gamma-shadow}
For every $\widehat F$ and every $\widehat\Lambda\in\gamma_{\mathrm{shadow}}(\widehat F)$,
\[
  \|\alpha_X(\widehat\Lambda)-\alpha_X(\Lambda)\|_2
  \leq
  \|\widehat F-\alpha_X(\Lambda)\|_2.
\]
Hence, on the event of \cref{cor:pgc-shadow},
\[
  \|\alpha_X(\widehat\Lambda)-\alpha_X(\Lambda)\|_\infty
  \leq\sqrt{|X|}\,\varepsilon.
\]
\end{corollary}
\begin{proof}
The objective in \cref{eq:gamma-sdp} is $\|\alpha_X(J)-\widehat F\|_2^2$. Since $\alpha_X(\Chan)$ is convex and compact, $\alpha_X(\widehat\Lambda)$ is the Euclidean projection of $\widehat F$ onto $\alpha_X(\Chan)$. Projection onto a closed convex set is non-expansive towards every point in the set, including $\alpha_X(\Lambda)$.
\end{proof}

Thus, $\gamma_{\mathrm{shadow}}$ selects a canonical CPTP representative among channels consistent with the measured fingerprint. This reconstruction is optional: fingerprint comparison itself operates directly on the $|X|$ entries and does not require a Choi matrix. The SDP acts on a $4^n\times4^n$ matrix and therefore has exponential size in $n$.

\subsection{Supplementary Validation of Theoretical Bounds}
\label{sec:app:bounds}
\label{sec:theory2Emp}

The main evaluation reports each result with its corresponding research question. The following figure (\autoref{fig:bounds}) gathers additional checks of the analytic bounds and estimator behaviour in one place.

\begin{figure}[t]
\centering
\includegraphics[width=0.98\linewidth]{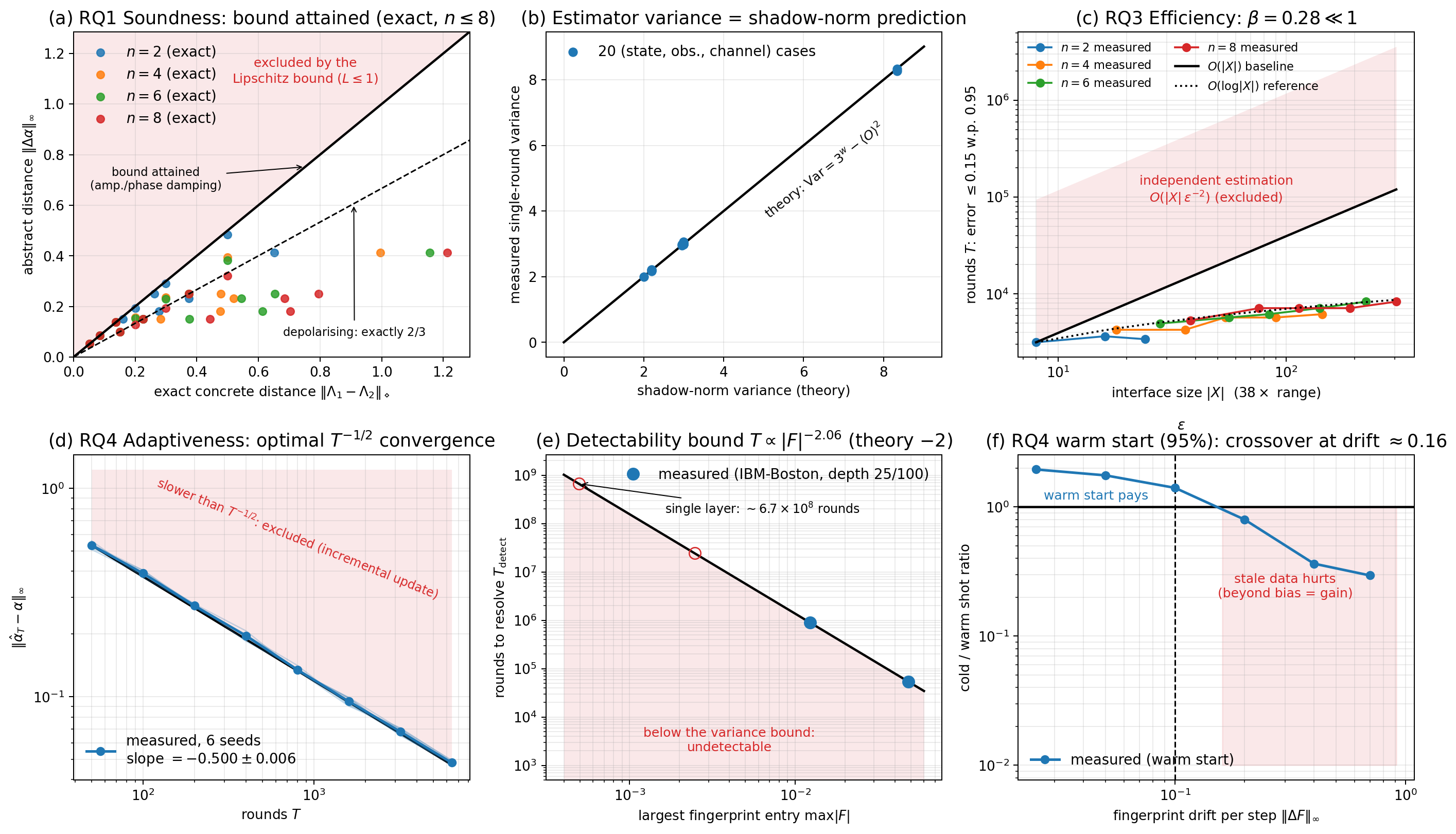}
\Description{Six panels comparing measurements with theoretical bounds: soundness relative to the unit-slope Lipschitz bound; measured variance against the shadow-norm identity; sample complexity as the interface grows to eight qubits; estimator convergence at the inverse-square-root rate; detectability cost against the inverse-square law; and the warm-start ratio crossing one as drift increases.}
\caption{Supplementary experimental checks of the theoretical bounds. Shaded regions are excluded by the corresponding bounds.}
\label{fig:bounds}
\end{figure}

\mypara{Soundness and variance}
Exact channel pairs at $n\leq8$ lie on or below the Lipschitz bound $L=1$, attained by amplitude- and phase-damping pairs; depolarising pairs attain exactly $2/3$. The soundness check includes three hardware-informed IBM-Boston single-qubit noise profiles (depolarising, amplitude damping, and phase damping applied to one qubit) and two homogeneous $n$-qubit product-noise profiles (depolarising and phase damping applied to every qubit). The measured single-round variance equals the shadow norm $3^w-\langle O\rangle^2$ to within a maximum deviation of $2\%$ over $20$ cases: the states $\ket{+}$, $\ket{0}$, and $\ket{+i}$ under depolarising noise ($p=0.1$) and amplitude damping ($\gamma=0.2$), measured with $X$, $Y$, and $Z$, and the Bell state $\ket{\Phi^+}$ under two-qubit depolarising noise ($p=0.1$), measured with $XX$ and $ZZ$.

\mypara{Sample complexity and estimator convergence}
Rounds required to reach $\varepsilon$ with confidence \ReachConf{} are measured against $|X|$ up to $n=8$. Shot budgets are searched on a multiplicative grid with factor $1.25$, and scaling exponents are estimated by least-squares regression over all measured points under homogeneous depolarising noise ($p=0.10$). The measured points lie far inside the linear-baseline region and near the logarithmic reference. Separately, six independent seeds track the predicted $T^{-1/2}$ convergence line with fitted slope $-0.500\pm0.006$.

\mypara{Detectability and drift boundaries}
The detectability threshold follows the $|F|^{-2}$ variance bound; extrapolating the measured hardware-rate results to a single layer gives approximately $\LoneTsingle$ rounds for the IBM-Boston profile. The warm-start benefit depends on the fingerprint displacement: when $\|\Delta F\|_\infty$ is small relative to the target accuracy $\varepsilon$, previous snapshots remain informative and reduce the fresh-measurement requirement. As drift grows, the bias from stale snapshots eventually dominates and the update no longer helps. The detailed hardware-rate and drift-limit experiments are reported in \cref{sec:res:limits}.

\subsection{Scaling the Evaluation}
\label{sec:scale}

We extend the empirical evaluation beyond the $n\le2$ limit imposed by exact diamond-norm SDPs in \Cref{sec:eval}. 

\begin{figure}[t]
    \centering
    \captionsetup{hypcap=false}
    \Description{Description of the growth figure.}
\includegraphics[width=0.5\linewidth]{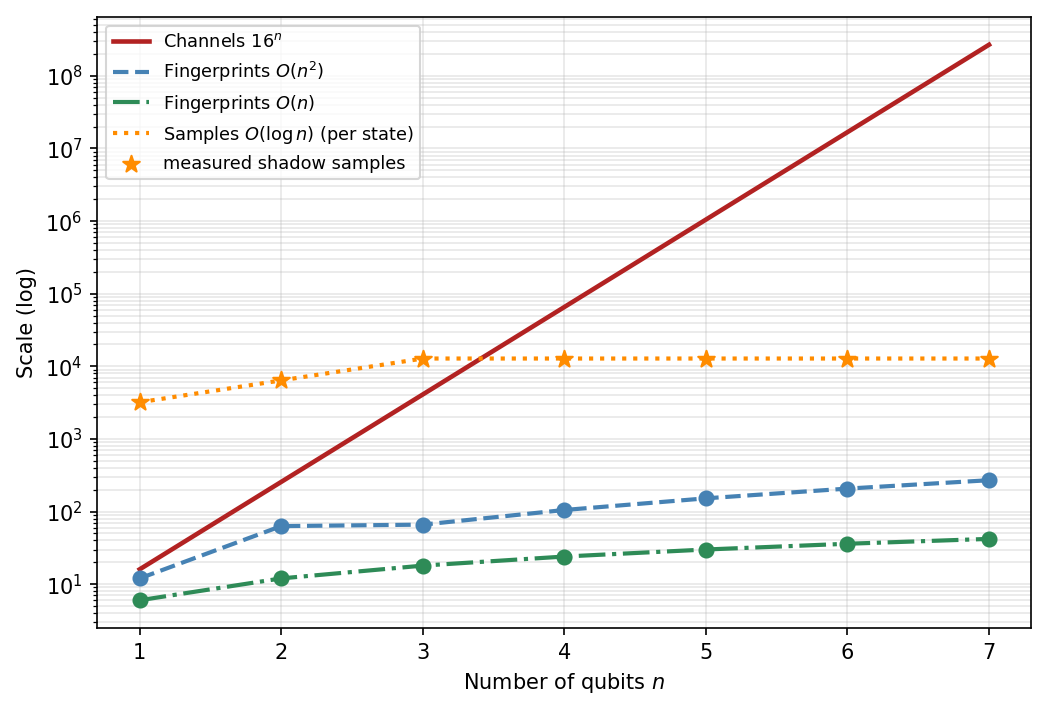}
    \Description{Log-scale plot over one to seven qubits
    showing the full channel dimension $16^n$, linear
    and quadratic fingerprint-size references, the
    logarithmic per-state sample-cost reference, and
    measured shadow-sample counts marked by stars.}
    
    \captionof{figure}{\textbf{Representation and
    measurement scaling.} Theoretical growth references
    for the full channel dimension ($16^n$), fingerprint
    size ($O(n)$ and $O(n^2)$), and per-state shadow-sample
    cost ($O(\log n)$), with measured shadow-sample
    counts for $n=1$--$7$.}
    \label{fig:asymptotics}
\end{figure}

\autoref{fig:asymptotics} contrasts the exponential growth of the full channel representation with the polynomial fingerprint-size references, and compares the predicted per-state sample-cost trend with measured shadow-sample counts.

Next, we describe the methodological extensions that enable scalable evaluation, and then report the corresponding evaluation results.

\subsubsection{Scaled Ground Truths and the Product Sampler}
\label{sec:method:scale}

Exact diamond norms via SDP are limited to $n \le 2$. We extend the exact evaluation to larger
channels with two ground truths, each validated against the SDP by the machine-checked test
suite before any experiment may use it: (i)~for channels identical on all but one site,
multiplicativity of the completely bounded trace norm gives
$\lVert \Gamma \otimes (\Lambda_1 - \Lambda_2)\rVert_\diamond =
\lVert \Lambda_1 - \Lambda_2\rVert_\diamond$, reducing the $n$-qubit distance to a
single-qubit SDP at any $n$; and (ii)~for Pauli channels, the diamond distance equals the
$\ell_1$ distance of the Pauli probability vectors, exact for depolarising and dephasing
products up to $n = 6$ here. On the estimation side, a product-factorised implementation of the
literal sampler draws each qubit's measurement outcome from its own rotated marginal---exactly
the distribution of \cref{alg:alpha-shadow} on product instances, verified against the full
sampler including $Y$ bases---at $O(nT)$ cost, enabling the genuine estimator at $n = 8$. The
scaled runs use product reference states and weight-$\le 2$ observables; entangled probes are
retained in the $n \le 2$ suites.

\subsubsection{Scaling the Evaluation: Exact Ground Truths at \texorpdfstring{$n \le 8$}{n <= 8}}
\label{sec:res:scale}

\begin{figure}[t]
\centering
\includegraphics[width=\linewidth]{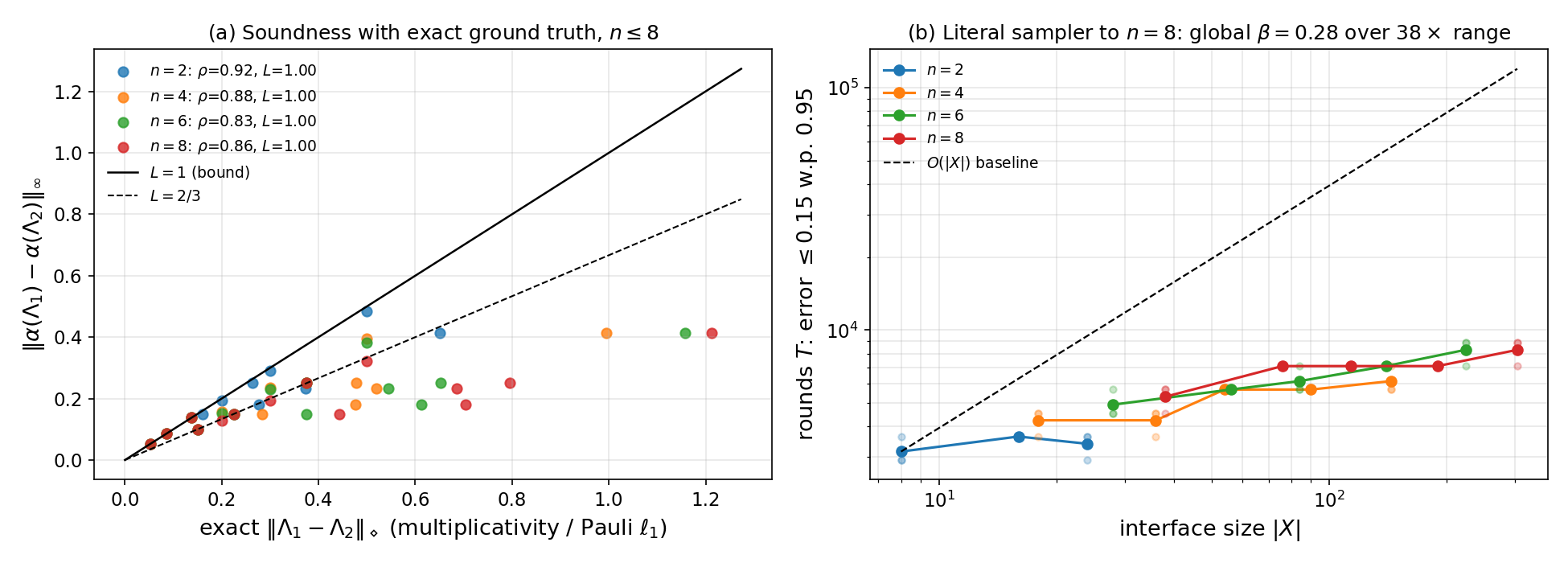}
\Description{Four panels: soundness scatter with exact ground truths at two, four, and six
qubits, points on the unit-slope bound and a two-thirds line with global families below; sample
complexity curves up to eight qubits far under the linear baseline; per-qubit rate recovery bars
all at one; and a telescoping reconstruction bound decreasing with budget.}
\caption{Scaling the evaluation with exact ground truths. (a)~Soundness at $n \le 8$: the bound
$L=1$ is attained at every size by single-site damping pairs; global Pauli families fall below
$2/3$ as $n$ grows. (b)~Literal-sampler rounds to $\varepsilon=0.15$ versus $\lvert X\rvert$ up
to $n=8$ ($\ScaleXrange\times$ range), pooled $\beta=\ScaleBeta$.}
\label{fig:scale}
\end{figure}

Diamond-norm SDPs limit exact evaluation to $n \le 2$; we lift this with two exact ground
truths, each machine-validated against the SDP before use (\cref{sec:method:scale}), and a
product-factorised literal sampler, and re-measure the research questions at scale
(\autoref{fig:scale}).

\mypara{Soundness (RQ1) at $n \le 8$}
Over $14$ exact channel pairs per size---single-site amplitude-damping, phase-damping, and
depolarising perturbations of an IBM-tier product channel (ground truth by cb-norm
multiplicativity) and global depolarising/dephasing Pauli products (ground truth by the Pauli
$\ell_1$ formula)---the Lipschitz bound is attained at \emph{every} size: $L = 1.000$ at
$n = 2, 4, 6$, realised by the single-site amplitude- and phase-damping pairs, with depolarising
fixed at exactly $2/3$. The correlation degrades gracefully ($\rho = 0.92 \to 0.83$), and the
degradation has a clean cause: \emph{global} Pauli families lose tightness with $n$ (worst ratio
$0.29$ at $n=6$) because their diamond worst case requires entangled inputs across many qubits,
which product probes do not supply. Soundness at scale is therefore tight exactly where the noise
is local---the regime the interface is designed for---and conservative for globally correlated
noise.

\mypara{Conciseness (RQ2) as per-qubit identifiability}
At $n = 4$ and $6$, regressing each qubit's amplitude-damping rate from the exact fingerprint
achieves $R^2 = 1.000$ for \emph{every} qubit, while regressions restricted to entries not
involving that qubit have $R^2 < 0$: the fingerprint's information is not merely retained but
\emph{localised} to the correct qubit, which is what fault attribution requires.

\mypara{Efficiency (RQ3) to $n = 8$}
With the literal product sampler, the rounds to uniform error $\varepsilon=0.15$ at confidence \ReachConf{} grow from $\ScaleTfirst$
to $\ScaleTlast$ while the interface grows $\ScaleXrange\times$ (from $\lvert X\rvert = \ScaleXmin$ to $\ScaleXmax$): a
pooled exponent $\beta = \ScaleBeta$ across sizes and seeds, and $\beta \approx \ScaleBetaWithinMin$--$\ScaleBetaWithinMax$ within
each size---far below the linear baseline over a range $5.7\times$ wider than the $n\le2$
measurement, materially strengthening \cref{def:est-eff} evidence.

\mypara{Adaptiveness (RQ4) via structured concretisation}
The general SDP concretisation is intractable beyond $n=2$ ($4^n$ Choi growth), so at $n = 4$ we
evaluate a structured product concretisation: per-qubit rates fitted from the estimated
fingerprint, with per-qubit reconstruction errors computed by \emph{exact} single-qubit diamond
norms and the total reported as the telescoping upper bound
$\sum_q \lVert\widehat\Lambda_q - \Lambda_q\rVert_\diamond$. The bound falls from $0.72$ to
$0.17$ as the budget grows $400 \to 6{,}400$ rounds---a bound, not an equality, and labelled as
such.

\subsection{Realistic Noise Framework and Stress Protocol: Probing the Limits}
\label{sec:method:limits}

The main evaluation uses canonical Kraus noise channels. 
We add a configurable per-qubit noise framework and four stress tests that probe where the concrete domain \emph{fails}. Hence, this section extends the evaluation towards hardware-inspired settings to characterise the conditions under which the abstraction remains applicable.

\subsubsection{Framework and Calibration}
Channels are assembled from per-qubit configurations---lists of (depolarising, amplitude-damping,
phase-damping) primitives at arbitrary rates---optionally composed with correlated two-qubit terms
(a CZ-attributed two-qubit depolarising channel and coherent $ZZ$ crosstalk
$e^{-i\theta ZZ/2}$) and repeated to depth $k$. The \texttt{ibm\_boston} profile (Heron r3,
156 qubits) uses the public platform calibration table (read 2026-06-10): two-qubit error median
$1.19\times10^{-3}$ (layered $2.28\times10^{-3}$) and readout error median $4.15\times10^{-3}$;
per-qubit $T_1, T_2$ and single-qubit gate error are Heron-class representative medians, replaceable
from the platform's downloadable calibration data. Gate-time/coherence ratios convert to per-gate
amplitude-damping and dephasing rates, the residual gate infidelity to a depolarising rate, and the
published per-pair spread fixes best/median/worst qubit tiers for heterogeneous configurations. The
calibrated readout error is injected into the literal estimator as a per-qubit assignment flip.

\subsubsection{Stress Tests}
\textbf{(L1) Detectability floor:} at hardware rates, how many rounds are needed to resolve the
largest fingerprint entry of a depth-$k$ layer to a factor-two signal-to-noise, and what bias floor
does readout error impose? \textbf{(L2) Non-Pauli errors:} is a coherent over-rotation (outside
every Pauli model) detected and recoverable by the general concretisation, and is the probe suite
first-order sensitive to coherent $ZZ$ crosstalk? \textbf{(L3) Blindness:} the probe map is a
finite set of linear functionals on the $16^n$-dimensional Choi space; we compute the dimension of
its null space for increasingly rich interfaces and \emph{construct} an explicit channel pair with
identical fingerprints, reporting its exact diamond distance as a measured lower bound on the
\emph{blindness radius}. \textbf{(L4) Drift validity region:} we sweep the drift magnitude and
locate the crossover at which pooling stale snapshots stops paying.

\subsubsection{Heterogeneous per-Qubit Noise}
To stress the protocol on realistic, spatially varying noise, we compose distinct per-qubit channels
and probe entangled states. The \emph{critical case} leaves one qubit noiseless and applies strong
amplitude damping ($\gamma=0.30$) to the other, then probes the Bell state $\ket{\Phi^+}$; a correct
fingerprint must attribute the fault to the damaged qubit (a large single-qubit deviation there, none
on the clean qubit) while still exposing its propagation through the correlator. We additionally test
nominally asymmetric rates representative of a calibrated device.

\subsubsection{Limits of the Concrete Domain}
\label{sec:res:limits}

The following four experiments identify the practical limitations of the abstraction (L1-L4). Each measures one aspect of the operating regime where the abstraction ceases to be reliable or informative.

\mypara{(L1) Detectability has a hard sample floor}
Resolving the largest fingerprint entry of a depth-$k$ IBM-Boston layer requires
$T \propto \lvert F\rvert^{-\LoneExpo}$ rounds at confidence \ReachConf{} over the measured decade (theory $-2$): $\LoneTdepthHundred$
rounds at depth $100$ ($\max\lvert F\rvert = \LoneFdepthHundred$) and $\LoneTdepthTwentyFive$ at depth
$25$, but an extrapolated $\sim\!\LoneTsingle$ for a \emph{single} layer
($\max\lvert F\rvert = \LoneFsingle$). Per-gate fingerprints at hardware rates are therefore
impractical; the operating regime is the depth-accumulated channel, where deviations compound
linearly with depth and the cost falls quadratically. The calibrated readout error
($4.15\times10^{-3}$) additionally imposes a bias floor of order $10^{-2}$ on the uncorrected
estimator: entries below the floor are invisible without readout mitigation, which is the practical resolution limit of the live protocol.

\mypara{(L2) Non-Pauli errors: detected and recoverable, with one blind spot}
A coherent over-rotation $R_X(0.1)$---outside every Pauli noise model---produces the full
first-order signature ($\max\lvert F\rvert = \sin\theta$), and the \emph{general} SDP
concretisation reconstructs the coherent channel, with diamond error falling $0.144 \to 0.035$ as the budget grows (the plateau is the SDP regularisation floor). Coherent $ZZ$ crosstalk, however,
exposes a structural blind spot: every observable in the paper suite commutes with $ZZ$ to first
order on the probe states, so the suite's response is $O(\theta^2)$
($\max\lvert\Delta F\rvert = 5.0\times10^{-3}$ at $\theta=0.1$), while a two-observable
extension ($YZ$, $ZY$ on $\ket{+0}, \ket{0+}$) restores the first-order signal
($\Delta F = \sin\theta = 0.0998$). Interface design is thus part of the soundness contract:
crosstalk-sensitive suites must include mixed-axis correlators.

\mypara{(L3) Blindness is real, measurable, and repairable}
The probe map of the paper's two-qubit suite spans only $80$ of the $256$ Choi dimensions, leaving
a $176$-dimensional blind subspace; we construct an explicit CPTP pair with
$\lVert\Delta F\rVert_\infty < 10^{-14}$ yet exact diamond distance $\LthreeDiamond$---a measured lower
bound on the blindness radius. This is information absent from the interface, not a failure of the
estimation guarantee of \cref{def:est-eff}. Enriching the interface shrinks the blind space monotonically: all
$15$ two-qubit Paulis leave $120$ dimensions; $16$ informationally complete product states with all
$15$ Paulis eliminate it entirely (blind dimension $0$), at which point $\gamma$ is exact in
principle. Conciseness and observational completeness trade off along this ladder, and the
fingerprint exposes the dial.

\mypara{(L4) Incremental updates have a validity region}
At confidence \ReachConf{}, warm-starting pays a factor $\LfourA\times$ at fingerprint drift $0.025$ per step and $\LfourB\times$ at
drift $\varepsilon = 0.10$, crosses over near drift $\LfourCross$, and \emph{hurts}
beyond it (ratio $\LfourC\times$ at drift $0.7$): stale snapshots carry bias equal to the drift, which
pooling dilutes only while it is below the statistical gain. These results empirically characterise the operating region of the incremental update procedure. They do not change \Cref{def:incremental}, which concerns estimator convergence under a fixed channel, nor do they establish a general guarantee that reusing stale measurements will improve accuracy. Rather, they show that the benefit of incremental updating depends on the magnitude of the fingerprint displacement and the resulting trade-off between sampling variance and bias.

\mypara{Custom profile}
The high-noise per-qubit profile (depolarising $5\times10^{-3}$, amplitude damping
$2\times10^{-3}$, phase damping $5\times10^{-3}$) runs through the full pipeline: exact
$\max\lvert F\rvert = 9.0\times10^{-3}$ concentrated on coherence entries (dephasing-dominated,
as expected), and the literal estimate at $T=2\times10^{5}$ followed by the general reconstruction
recovers the channel to diamond error $5.0\times10^{-3}$---the framework is usable end-to-end at
realistic rates given the L1 budget.


\end{document}